%% file: main.tex
\documentclass[11pt,letterpaper]{article}
\usepackage{dynamicm-eprint}
\usepackage{amsmath,amssymb,amsthm,mathtools}
\usepackage{booktabs,longtable,tabularx,array}
\usepackage{enumitem}
\usepackage{xcolor}
\usepackage{tikz}
\usetikzlibrary{arrows,arrows.meta,positioning,calc}
   
\definecolor{navy}{RGB}{28,55,82}
\definecolor{slate}{RGB}{78,88,99}
\definecolor{pale}{RGB}{242,246,249}
\definecolor{proved}{RGB}{25,105,65}
\definecolor{greenstatus}{RGB}{26,105,65}
\definecolor{conditional}{RGB}{128,63,84}
\definecolor{tentative}{RGB}{135,52,65} 
\definecolor{open}{RGB}{165,93,20}
\definecolor{openstatus}{RGB}{164,92,18}
\definecolor{orangestatus}{RGB}{156,88,18}
\definecolor{redstatus}{RGB}{145,45,45}
\definecolor{refuted}{RGB}{150,43,43}

\setlist[itemize]{leftmargin=1.4em,itemsep=2pt,topsep=3pt}
\setlist[enumerate]{leftmargin=1.65em,itemsep=2pt,topsep=3pt}

\ifLuaTeX
  \usepackage{selnolig}  
\fi
\IfFileExists{bookmark.sty}{\usepackage{bookmark}}{\usepackage{hyperref}}
\IfFileExists{xurl.sty}{\usepackage{xurl}}{} 
\hypersetup{
  pdftitle={Opening the Strategic Pandora Box: Conditional Transaction Mechanisms},
  pdfauthor={Yonatan Sompolinsky, Ori Newman, Xintong Wang, David Parkes},
  colorlinks=true,
  linkcolor={dynamicmlink},
  filecolor={Maroon},
  citecolor={dynamicmlink},
  urlcolor={dynamicmlink},
  pdfcreator={LaTeX}}

\title{Opening the Strategic Pandora Box: Conditional Transaction Mechanisms}
\author{%
  Yonatan Sompolinsky$^{1}$ \qquad Ori Newman$^{2}$\\[0.75ex]
  Xintong Wang$^{3}$ \qquad David Parkes$^{1}$\\[1ex]
  \small
  $^{1}$Computer Science, Harvard John A. Paulson School of Engineering
  and Applied Sciences, Harvard University\\
  $^{2}$Blavatnik School of Computer Science and AI, Tel Aviv University\\
  $^{3}$Department of Computer Science, Rutgers University
}
\date{}

\begin{document}
\input{00-title-and-abstract.tex}
\input{01-introduction.tex}
\input{02-model-and-rw.tex}
\input{03-restricted-nbir.tex}
\input{04-known-values.tex}
\input{05-reported-values.tex}
\input{07-economic-state-machine.tex}
\input{acknowledgments.tex}
\input{references.tex}
\clearpage
\pdfbookmark[0]{Appendices}{appendices-root}
\section*{Appendices}
\appendix

\newcommand{\appendixsection}[2]{%
  \refstepcounter{section}%
  \section*{Appendix~\thesection: #1}%
  \addcontentsline{toc}{section}{Appendix \thesection: #1}%
  \label{#2}}

\input{appendices/appendix-a-discounted-pandora-ordering.tex}
\input{appendices/appendix-b-restricted-nbir.tex}
\input{appendices/appendix-c-pure-rw-revenue.tex}
\input{appendices/appendix-d-ordinary-mixed-rw-revenue.tex}
\input{appendices/appendix-e-two-box-separation.tex}
\input{appendices/appendix-f-reported-values-membership.tex}
\input{appendices/appendix-g-reported-values-mixed-revenue.tex}
\input{appendices/appendix-h-reported-values-constant-competition.tex}
\input{appendices/appendix-i-existence-boundary.tex}
\input{appendices/appendix-j-open-domain-nonexistence.tex}
\end{document}

%% file: 00-title-and-abstract.tex
\maketitle

\section*{Abstract}\label{abstract}
\addcontentsline{toc}{section}{Abstract}

Conditional transaction engines (CTEs) execute conditional instructions for offline
users. This paper formalizes the mechanism-design problem within each engine invocation.
A conditional transaction mechanism (CTM) decides which pending conditions to evaluate
first because each evaluation delays the eventual write. We model this problem as
Strategic Pandora, a discounted variant of the Pandora's box model with independent
Bernoulli boxes. Agents report privately assessed success probabilities and, in the full
model, values for the write action. We propose the reported-Weitzman mechanism (RW) and
the reported-values second-price mechanism (RWSP). To compare revenue without a common
prior, we introduce dynamic No-Betting Revenue. Under the stated competition and
equilibrium conditions, every qualifying pure equilibrium of RW or RWSP earns a constant
fraction of its corresponding dynamic NBR benchmark.

%% file: 01-introduction.tex
\section{Introduction}\label{introduction}

On October 10, 2025, an estimated \$3.21 billion in leveraged positions in crypto was liquidated
in a single minute at the peak of a market cascade. Reports at the time put total
liquidations at more than \$19 billion across roughly 1.6 million trading accounts.
In November 2022, FTX customers became unable to withdraw their assets when the
exchange halted withdrawals. In a later submission to the CFTC, Better Markets
raised the general concern that automatic liquidation in markets that operate
around the clock can catch users off guard or asleep as liquidation thresholds
approach (Marshall 2025; Axios 2025; CFTC 2022; Better Markets 2025).

\input{figures/esm-motivation.tex}

Running a 24/7 monitoring server is unsustainable for non-expert traders and family
offices. In that, our framework is appealing for the long tail of users who cannot run a
24/7 monitoring server. Moreover, servers can fail during network stress and therefore
do not guarantee access during sudden shocks such as the October cascade. The framework
is therefore also useful for expert users and entities whose monitoring servers may
lose access when the entire system is stressed. The FTX collapse involved a
platform-level failure that our framework does not address, but it shows that access
to a centralized platform cannot be assumed. These cases motivate protection that can
act on behalf of offline users without requiring them to transmit new instructions in
real time.

We consider conditional transaction engines (CTEs), which evaluate and sequence
instructions while their users are offline. Users submit ``If This Then That'' (IFTTT)
instructions in advance. At each invocation, the CTE invokes a conditional transaction
mechanism (CTM). The CTM uses reported condition probabilities and write values to
determine which conditions to evaluate and which instruction receives the next
transaction slot. This architecture extends the
capability of existing transaction engines, such as Ethereum's EVM, which sequence
transactions in real time only and cannot trigger transactions autonomously. CTEs remain relevant
even when the system is run by one logical server or sequencer, as
at a centralized exchange such as Nasdaq. However, as we explain, the primary use case
is expressive composable systems such as DeFi, where transactions can interact
atomically with multiple markets or logic zones. Composite transactions render Price
First Time Second and similar TradFi policies unsuitable, since no single price can rank a composite rule. 
Submission time too is irrelevant, because rules may be submitted days ahead of their triggering.

An empirical study of Aave liquidations on October 10 documents speculative
traders repeatedly submitting transactions in anticipation of liquidations. One Aave liquidator submitted 
14,333 transactions in an attempt to time the event, only to have 99.05\% of those transactions reverted 
(Sevim and Ferreira Torres 2026, Section~6.2 and Table~9). 
Our framework obviates such laborious submission processes: The user submits one persistent condition or action rule, and the rule remains available to the sequencer
across invocations. The sequencer evaluates its condition and, when the
condition is satisfied and the rule's user is selected in the auction, enacts on the user's behalf with theoretically
zero latency. Furthermore, when the condition holds but some other participant outbids the user, the mechanism avoids
charging the latter, relieving the costs of participation.

In fact, the very need for liquidators stems from smart contracts' inability to trigger their own conditionals. 
Lacking such capabilities, DeFi lending contracts incentivize and rely on third parties to listen to 
on-chain state (e.g., an oracle update on Chainlink contract) and trigger liquidation when needed
(Aave n.d.). Our paradigm alleviates such workarounds: 
Using a CTE, the maintainers can encode in advance the entire
liquidation logic as a conditional instruction,
and avoid manual liquidators altogether.\footnote{For liquidation to be applied atomically
with its triggering condition, the maintainors should attach sufficient bids.
In our paradigm, they can attach bids for 'write' only, and rely on the reports of online 'read' agents to
increase the 'read' bids according to real-time pressure.}

\subsection{Existing approaches}

Delegated execution is not new. Brokers in TradFi and keepers and solvers in DeFi monitor specified conditions and
submit trades or transactions when those conditions hold (Gelato Network 2023;
Chainlink 2023; Anoma 2024). More generally, subscription and listener interfaces,
event-driven programming systems, IFTTT-style automation, and platforms such as Reactive
Network evaluate predicates and dispatch actions (Kaleem et al.\ 2021;
Zhao et al.\ 2022; Reactive Network 2024). However, these services operate outside the
sequencing engine that commits the final state transition. They can attempt submission,
but they do not determine admission or priority. The mechanism-design problem therefore
remains: how should the sequencing engine allocate condition reads and the next
transaction slot among pending rules?

Event-based smart contract systems were proposed in the past, albeit absent a mechanism-design-informed architecture. These systems
largely treat triggered transactions as non-mandatory inputs for the sequencer, and do not 
treat evaluation inside their systems. This renders those systems an incremental improvement over existing smart contract systems.
 In contrast, a CTE can define deterministic mandatory prioritization that operates
 in-consensus, regardless of the operator's cooperation. Its CTM treats reads and writes
 in the same mechanism.


\subsection{The one-shot allocation problem}
\label{the-strategic-pandora-box}

Evaluating a rule's condition and altering the shared state constitute two distinct
resource-allocation problems. The former is a 'read', and as such it is commutative: it consumes computation,
but does not inherently preclude evaluation of other rules, and the ordering has no consequence.  
Executing a satisfied rule, in contrast, 
alters the shared state and does not generally commute with other writes. The mechanism allocates at most
one rule to the next transaction slot.

Evaluating too many conditions still imposes a prohibitive externality on the system, since each read delays
further the eventual write. We model the process through a discount factor
applied to the value of the write.

Coupling thus the reads and writes allows for more efficient mechanisms that avoid spending read cycles 
if the corresponding write is outbid by bidders on other queries.
For instance, conditional instructions of the Aave liquidator
 discussed above will not be read and evaluated if the liquidator is not guaranteed to receive priority on the write,
 as it is still outbid by counterparts.

The resulting problem maps into a discounted variant of the Pandora's box problem with independent Bernoulli boxes,
where each pending rule is represented by a box with unknown parameters. Crucially, the system is dynamic: rules' success probabilities
may change with the state and environment, and agents may continuously revise their bids accordingly.
Our paper focuses on the
allocation of reads and writes at a given point in time. We leave the broader dynamic
process for separate work.

In our multi-agent version of Pandora, agents report the \texttt{True}
probability of pending rules they are interested in. Agents may hold very different beliefs, 
and the mechanism remains prior-free and does not aim to learn the real underlying parameter.
\footnote{This separates our framework from traditional exploration-exploitation problems, and from 
no-regret framework. This is motivated by the environment being noisy and dynamically changing, and by 
one primary use case being protection from market catastrophes which is inherently not learnable.} 

In the full \emph{reported values} model, agents report also their value or bid for the write action.  
However, for pedgagoical reasons, we study also an interim model, the \emph{known box values},
where box values are fixed and known, and agents report read-probabilities only.
 

For the one-shot game, we introduce two mechanisms, one for each model:
the reported-Weitzman mechanism (RW) and the reported-values second-price mechanism
(RWSP).
Both mechanisms adapt Weitzman's optimal solution to the planner's Pandora problem to
a mechanism-design setting in which the mechanism must choose a payment vector. They
differ in how payments are discounted and in whether the write action is allocated as
in a first-price or second-price auction. Section~\ref{the-reported-weitzman-mechanism-1}
formalizes both mechanisms and explains these differences. We prove that each mechanism
obtains a constant fraction of the optimal mechanism in the corresponding model. The benchmark
restricts outcomes to an admissibility class which we now discuss.



\subsection{Prior-free mechanism design}
Mechanism design typically strives to maximize social welfare. However, when agents
hold different beliefs about the underlying uncertainty, and the mechanism itself is restricted
to remain prior-free, social welfare is undefined. Summing agents' subjective
utilities leads to betting, mnamely, to allocations that count ex-ante value
from transfers between agents holding opposing beliefs on the likelihood of events,
rather than from exogenous utility.    

Drawing inspiration from Gilboa et al.'s No-Betting-Pareto Dominance (NBPD) criterion, we introduce 
No-Betting Revenue (NBR) and dynamic NBR (dNBR) to define an admissibility class of outcomes. These 
criteria ensure that the framework does not reward mechanisms that obtain an artificially high revenue 
due to betting. We formalize the admissibility class in Section~\ref{belief-compatible-revenue-and-the-prior-free-revenue-benchmark},
and show why the NBPD criterion is inadequate for a mechanism design setup.

\subsection{Our contributions}\label{our-contributions}

\begin{itemize}
      \item We define the CTM, the one-invocation framework for conditional delegated
            execution. A CTM takes the active conditional rules, reports, and bids;
            determines which conditions to evaluate and in what order; assigns payments;
            and selects the next write. We show how this framework maps into the Strategic
            Pandora game.

      \item We introduce a new prior-free criterion for mechanism design with
            heterogeneous beliefs: NBR for a single decision and dynamic NBR for sequential
            decisions. NBR includes the auctioneer in the comparison and ranks admitted
            outcomes by deterministic revenue. Dynamic NBR uses range-constrained product
            beliefs to account for the value generated at sequential decisions. This
            contribution may be of independent interest.

      \item We devise the RW mechanism for known box values and the RWSP mechanism for
            reported values. Under the stated competition and equilibrium conditions,
            every qualifying pure equilibrium of each mechanism earns a constant fraction
            of the corresponding dynamic NBR revenue benchmark.
\end{itemize}

\subsection{Conditional Transaction Engines}

The transaction seletion mecahnism, the CTM, is only one component within the broader engine, the CTE.
The former is responsible, at the end of every transction slot, to process a certain snapshop of 
conditional rules and their bids, and output the active rule(s) to evaluate,
and to allocate the next transaction slot.

The CTE handles the dynamic process of running a series of CTM instances, one after each transaction slot.
The CTE should ideally comprise other protocols too, in particular ones governing bid submisson and
revision. A properly devised CTE protocol would ensure that the operator cannot engage in bid sniping
or free-ride the probability-agents' information.

A hollistic discussion of the CTE,
and of the operator's restricted role in real-time admission, will appear in a separate work. For our scope,
one can safely assume away the opeartor altogether, and focus on the question of choosing reads and write based
on a fixed input of conditional rules and their bids. The reader should further think of real-time transactions
as special trivial conditional rules of the form If \texttt{True} Then That.  


%% file: figures/esm-motivation.tex
\begin{figure*}[t]
    \centering
    \includegraphics[width=\textwidth]{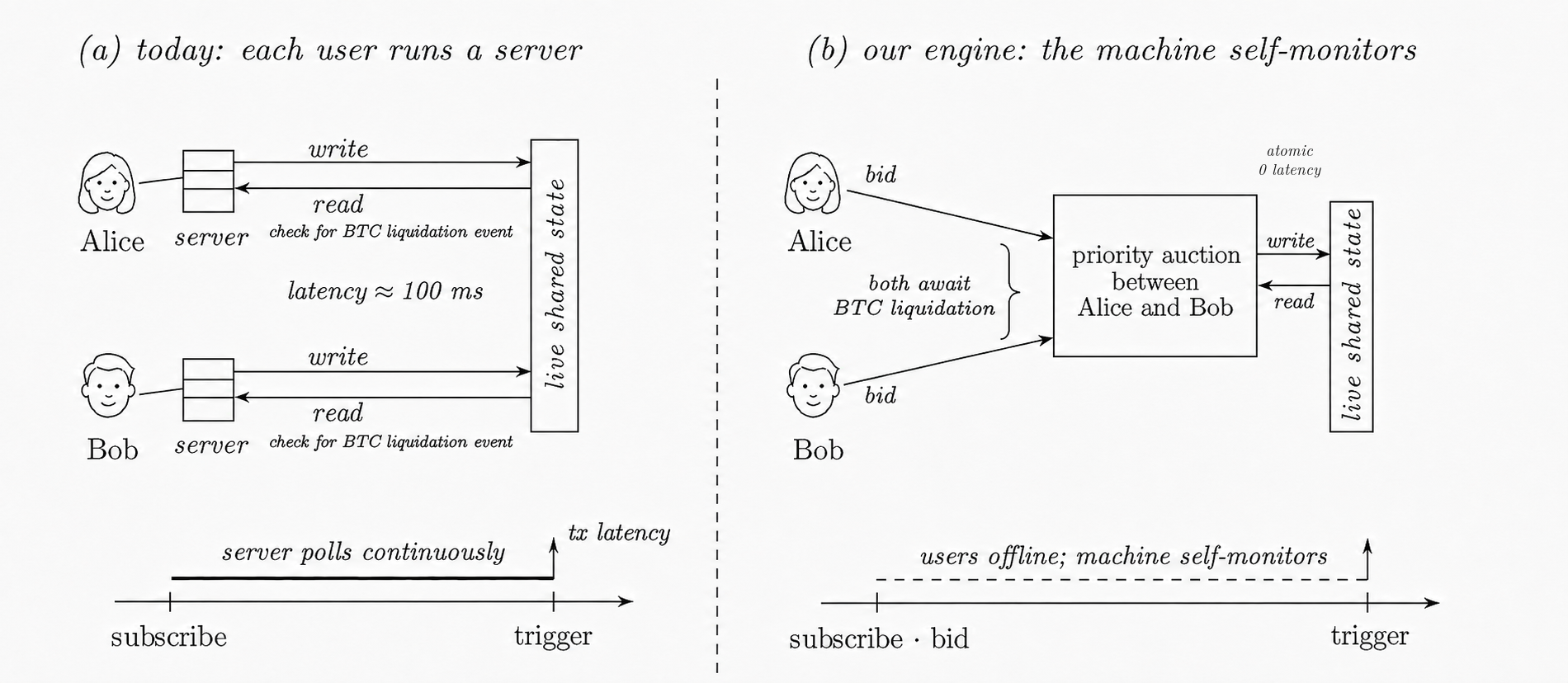}
    \caption{Two ways to act on a BTC liquidation event. (a) Alice and Bob each
    maintain an online process that reads the shared state and later submits a
    transaction. (b) Alice and Bob submit persistent value and priority information
    before the event; the engine evaluates the recorded conditions and selects among
    the resulting eligible transactions. The ``0 latency'' label denotes removal of an
    additional user-side reaction step, not zero physical or network latency. The
    comparison presumes neither admission nor execution.}
    \label{fig:motivation}
\end{figure*}

%% file: 02-model-and-rw.tex
\section{Strategic Pandora and RW}\label{the-discounted-pandora-model-and-the-rw-mechanism}

After each trasaction slot executes, the CTM chooses the next rules (or regular If \texttt{True} Then That tranasction)
to evaluate and own the next write slot. It does so based on a fixed set of read (probaboltiy) and write (value)
 bids. As mentioned above, the problem of the CTM designer can be formalized as a prior-free
 variant of the Pandora box problem.
 
\subsection{Discounted Pandora with Bernoulli boxes}\label{discounted-pandora-with-bernoulli-boxes}

There are \(M\) independent boxes. Box \(j\) contains \(v_j>0\) with probability
\(p_j\) and zero otherwise. Inspection is free but creates delay: a common
\(0<\delta<1\) discounts the box in position \(t\) by \(\delta^{t-1}\), and search
ends at the first success. For an order \(\pi=(\pi_1,\ldots,\pi_M)\), its value is
\[
      W_\delta(p,v;\pi)=
      \sum_{t=1}^{M}\delta^{t-1}
      \Bigl(\prod_{s<t}(1-p_{\pi_s})\Bigr)p_{\pi_t}v_{\pi_t},
      \qquad W_\delta(p,v)=\max_\pi W_\delta(p,v;\pi).
\]
The discounted Weitzman index is
\(z(p,v)=pv/(1-\delta+\delta p)\); box \(j\) has index
\(z_j=z(p_j,v_j)\).

The following ordering lemma determines the nonstrategic benchmark used throughout.

\textbf{Lemma 1 (Discounted Weitzman optimality).}
\emph{Among policies that execute the first successful box and then stop,
      ordering the boxes in weakly decreasing order of \(z_j\) maximizes \(W_\delta\).}

Appendix~\ref{app:discounted-pandora} proves the lemma by comparing adjacent boxes
and repeatedly swapping inverted pairs.

\subsection{Agents and reports}\label{agents-beliefs-and-reports}

There are \(n\) agents. Agent \(i\) privately assigns box \(j\) success probability
\(p_i^j\in(0,1)\), independently across boxes.\footnote{The restriction to interior
      true beliefs is imposed to exclude boundary cases immaterial to the main comparison
      results.} Under known box values, \(v_j\) is public
and the agent reports \(q_i^j\in[0,1]\). Under reported values, its true type is
\((p_i^j,v_i^j)\) and it reports \((q_i^j,b_i^j)\), with \(b_i^j\ge0\). For every box,
write \(\widehat q_j=\max_iq_i^j\) and let \(r(j)\) be its selected highest probability
reporter. In RW under known box values, the selected reader also receives the box's
value if that box is the first success. Under reported values, reader and writer roles
are selected separately as specified below.

\subsection{The two mechanisms}\label{the-reported-weitzman-mechanism-1}
\phantomsection\label{reported-mechanism}

\noindent
\fbox{\begin{minipage}[t]{0.455\textwidth}
            \vspace{0pt}\small
            \textbf{Mechanism 1: RW under known box values}\\[2pt]
            \textbf{Input:} Public values \(v_j\); reports \(q_i^j\).\\
            \textbf{Output:} An inspection order, up-front payments, and one executed box.
            \begin{enumerate}[leftmargin=1.25em,itemsep=1pt,topsep=3pt]
                  \item \textbf{Adopt reports.} Set \(\widehat q_j=\max_iq_i^j\) and select a reader
                        \(r(j)\) attaining the maximum.
                  \item \textbf{Rank.} Set
                        \(\widehat z_j=\widehat q_jv_j/(1-\delta+\delta\widehat q_j)\) and sort decreasingly.
                  \item \textbf{Charge.} At position \(t\), set
                        \(D_j=\delta^{t-1}\prod_{s<t}(1-\widehat q_{\pi_s})\) and charge the reader
                        \(P_j=D_j\widehat q_jv_j\) up front.
                  \item \textbf{Run.} Inspect in order and execute the first success.
            \end{enumerate}
      \end{minipage}}
\hfill
\fbox{\begin{minipage}[t]{0.455\textwidth}
            \vspace{0pt}\small
            \textbf{Mechanism 2: RWSP}\\[2pt]
            \textbf{Input:} Reports \((q_i^j,b_i^j)\).\\
            \textbf{Output:} An inspection order, reader and writer roles, payments, and one
            executed box.
            \begin{enumerate}[leftmargin=1.25em,itemsep=1pt,topsep=3pt]
                  \item \textbf{Select roles.} Set \(\widehat q_j=\max_iq_i^j\), reader \(r(j)\) at
                        that maximum, writer \(w(j)\) at the highest value report, and second price
                        \(y_j=b_j^{(2)}\).
                  \item \textbf{Rank.} Set
                        \(\widehat z_j=\widehat q_jy_j/(1-\delta+\delta\widehat q_j)\) and sort decreasingly.
                  \item \textbf{Charge.} Set
                        \(D_{r(j),j}=\delta^{t(j)-1}\prod_{k\prec j}(1-q_{r(j)}^k)\) and charge the
                        reader \(D_{r(j),j}q_{r(j)}^jy_j\) up front.
                  \item \textbf{Run.} Inspect in order. At the first success, execute the box and
                        transfer \(y_j\) from writer to reader.
            \end{enumerate}
      \end{minipage}}

\medskip
In RWSP, \(b_j^{(2)}\) is the second-highest value
report among distinct agents, with ties occupying separate ranks. A value tie is broken
by probability report, a probability tie by value report, and remaining exact ties by a
fixed deterministic rule. RW exact ties are also resolved deterministically. The
technical equilibrium-selection conditions used by the known-values results are stated
in Appendices~\ref{app:bcr} and~\ref{app:known-pure}.

\paragraph{The up-front objective-term payment rule.}
RW replaces the probability inputs to the discounted
Weitzman procedure with the most optimistic reports and attaches payments to the
resulting allocation rule. For each box, the
selected highest probability reporter pays up front the box's discounted contribution
to the Weitzman objective. Raising a probability report can
win the reader role and move the box earlier, but it also raises the payment. Lowering a
report reduces the payment but risks losing the role to a rival. Because the entire
objective term is charged before inspection, realized successes and failures do not
change mechanism revenue. Revenue equals the maximum discounted value under the most
optimistic report for each box:
\[
      \operatorname{Rev}_{\mathrm{RW}}(q)
      =\sum_jP_j
      =W_\delta(\widehat q,v).
\]

\paragraph{A two-box instance: order, delay, and payment.}
The following calculation applies RW to two portfolio-protection rules. It shows why
the highest success probability need not be inspected first, and how the same reports
determine what each selected reader pays before inspection. Take public write values
\(v_A=10\), \(v_B=6\), discount \(\delta=0.9\), and highest probability reports
\(\widehat q_A=0.5\), \(\widehat q_B=0.8\). Substituting into RW gives:
\begin{center}
      \small
      \begin{tabular}{@{}lcc@{}}
            \toprule
                                                    & Rule \(A\)            & Rule \(B\)              \\
            \midrule
            Public value \(v_j\)                    & \(10\)                & \(6\)                   \\
            Adopted report \(\widehat q_j\)         & \(0.5\)               & \(0.8\)                 \\
            Index \(z(\widehat q_j,v_j)\)           & \(5/0.55\approx9.09\) & \(4.8/0.82\approx5.85\) \\
            Inspection position                     & First                 & Second, if \(A\) fails  \\
            Discount times reported reach \(D_j\)   & \(1\)                 & \(0.9(1-0.5)=0.45\)     \\
            Up-front payment \(D_j\widehat q_jv_j\) & \(5\)                 & \(0.45(0.8)(6)=2.16\)   \\
            \bottomrule
      \end{tabular}
\end{center}
RW inspects \(A\) first because its index is larger. If \(A\) succeeds, it receives
the next transaction slot and \(B\) is never inspected. If \(A\) fails, \(B\) is
inspected one period later, so its successful write has discounted value
\(0.9\cdot6=5.4\). Both payments have already been collected, including the payment
for \(B\) when it is never reached; retained revenue is \(5+2.16=7.16\) in every
outcome. The calculation applies to the stated reports, without assuming equilibrium.

\paragraph{Separating probability reports from value reports.}
A direct first-price extension would combine the reader's
probability report with the writer's value report, charge the reader the box's
objective term, and charge the writer its own report after success. Pure equilibrium
may then permit writer overbidding, so the success transfer can exceed the selected
writer's true value and the dynamic NBR membership argument fails. RWSP gives execution
upon success to the writer with the highest
value report and computes the box's index using the second-highest reported value. This second-highest price also determines the writer's payment to the reader upon
a successful read. In the value-only, single-box specialization stated below, truthful
value reporting is weakly dominant. The discount on the reader's up-front payment is
calculated using that agent's reports about the previous boxes, rather than the adopted
maximum reports. An ex-ante charge to the writer is not a substitute: the writer may assign very low
probability to the condition while placing very high value on priority if it
occurs.\footnote{For example, a large liquidity provider may believe that the market is
      stable but place a high conditional value on priority in a black-swan crash.}

\medskip
\noindent\emph{Remark (second-price truthfulness and offline value submission).}
Consider a user protecting one financial position in a particular contract or liquidity
pool. For simplicity, assume that the user is single-minded: it has positive value for
the corresponding protection rule and zero value for all other boxes or events. Assume
also that it does not compete for the reader role. Holding the probability side and the
other value reports fixed, writer allocation and payment for this box are those of a
standard second-price auction. Truthful value reporting is therefore weakly dominant.
The value player may be a trader or high-net-worth investor rather than a server operator.
It reports only how much execution is worth to it; it need not estimate or monitor the
probability that the condition holds. In the application, the division is temporal as
well as informational: probability players are typically online and supply current
probability reports, whereas the value player may have less information about real-time
market conditions and remain offline. Probability--value decoupling therefore permits
the value player to benefit from the information supplied by these probability players.
It can submit its value and its execution and payment authorizations in advance without
operating a server continuously or communicating a fresh instruction during the event.
Once these inputs
are admitted to the mechanism state, going offline or losing real-time access creates
no additional loss of allocation or utility by itself.

Applied to October 10, this specialization removes two requirements from portfolio
protection: the value owner need not remain awake, and a previously admitted instruction
need not be transmitted when its condition holds. The statement concerns
participation by an already submitted rule, not price formation. It is conditional on
automatic execution of the recorded mechanism logic and does not guarantee admission,
platform liveness, execution, or that a stored value remains current.

The next section develops NBR for one decision and then dynamic NBR for sequential decisions.
Section~\ref{known-values-the-revenue-guarantee} then gives
the membership and revenue guarantees under known box values and reported values.

%% file: 03-restricted-nbir.tex
\section{No-Betting Pareto Dominance, No-Betting Revenue, and Dynamic No-Betting Revenue}
\label{belief-compatible-revenue-and-the-prior-free-revenue-benchmark}

The mechanisms in Section~\ref{the-discounted-pandora-model-and-the-rw-mechanism}
collect up-front payments from agents who may disagree about success probabilities.
This disagreement also affects the revenue benchmark. Contingent transfers can cover
each agent's up-front payment in expectation under that agent's own belief, even when
the mechanism allocates no value. Scaling these transfers then produces arbitrarily
large revenue. Individual rationality alone therefore gives no meaningful revenue
comparison for RW or RWSP.

The example below isolates the problem for a single decision. NBR excludes outcomes
whose revenue is supported only by incompatible beliefs. In Strategic Pandora, agents
may also disagree about the probability that a later box is reached. Dynamic NBR
applies the same restriction to the sequence of inspections.

\subsection{Example: revenue without allocated value}\label{example-disagreement-bet}

\paragraph{The decisions and transfers.}
There are two agents and two independent binary conditions \(A,B\). The mechanism
observes \(A\) first and observes \(B\) only if \(A\) fails; it stops at the first
success. The discount is \(\delta=1/2\). Before either observation, each agent pays
\(\Lambda>0\) to the mechanism. Subsequent transfers are between the agents, as follows:
\begin{center}
  \small
  \begin{tabularx}{\linewidth}{@{}lXc@{}}
    \toprule
    Observed event              & Transfer between agents                  & Discount \\
    \midrule
    \(A\) succeeds              & Agent 2 pays \(20\Lambda/13\) to agent 1 & \(1\)    \\
    \(A\) fails, \(B\) succeeds & Agent 1 pays \(64\Lambda/13\) to agent 2 & \(1/2\)  \\
    Both fail                   & No transfer                              & ---      \\
    \bottomrule
  \end{tabularx}
\end{center}
The mechanism retains the two up-front payments in every outcome. The two agents have
opposed interior beliefs:
\[
  (p_1^A,p_1^B)=(3/4,1/4),
  \qquad
  (p_2^A,p_2^B)=(1/4,3/4).
\]

\paragraph{Own-belief individual rationality.}
Each agent evaluates the discounted amount it receives minus the discounted amount it
pays to the other agent. The probability of the second transfer is the probability that
\(A\) fails times the probability that \(B\) succeeds. Under their respective beliefs,
these expectations are
\[
  \begin{aligned}
    \text{Agent 1:}\quad &
    \frac34\frac{20\Lambda}{13}
    -\frac12\frac14\frac14\frac{64\Lambda}{13}=\Lambda, \\
    \text{Agent 2:}\quad &
    -\frac14\frac{20\Lambda}{13}
    +\frac12\frac34\frac34\frac{64\Lambda}{13}=\Lambda.
  \end{aligned}
\]
Subtracting each agent's up-front payment \(\Lambda\) gives zero expected utility.
Thus own-belief individual rationality permits retained revenue \(2\Lambda\) for every
\(\Lambda>0\), although no value is allocated.

\paragraph{Excluding revenue without allocated value.}
Each agent regards the transfer it receives as sufficiently likely to cover its
up-front payment. Under any one belief used to evaluate both agents, however, their
transfers sum to zero at each decision. The revenue comparison must therefore require
support from allocated value in addition to own-belief individual rationality.

\subsection{NBR: revenue comparison for a single decision}\label{the-nbr-criterion}

In ``No-Betting-Pareto Dominance,'' Gilboa, Samuelson, and Schmeidler
(2014, Definitions 2--3) require every affected agent to gain strictly, both under
its own belief and under a single hypothetical belief. We adapt their comparison to
mechanism revenue relative to no action.

\paragraph{Include the auctioneer and allow zero agent surplus.}
Competition can make each agent's payment equal its expected gross payoff under the
belief used for the comparison, leaving every agent with zero expected surplus. A
revenue benchmark should admit such outcomes. We therefore include the auctioneer in
the comparison with no action. Positive revenue can supply the required strict gain
while every agent is weakly better off.

Fix one decision with finite uncertain outcome \(X\). Agent \(i\) has belief
\(P_i\), gross payoff \(U_i(X)\), and up-front payment \(t_i\). Gross payoff
includes any contingent transfers between agents, which balance in every state.
Write \(u_i(X)=U_i(X)-t_i\). Let the auctioneer be participant \(0\), with
\[
  u_0(X)=R=\sum_{i=1}^n t_i.
\]
Revenue is deterministic: for the mechanism outcome being evaluated, its amount does
not vary with the realized state. The auctioneer supplies no additional belief or
strategic report. Normalize every participant's payoff under no action to zero, and
retain weak own-belief individual rationality,
\(\mathbb E_{P_i}[u_i]\ge0\) for each agent \(i\).

The comparison requires one certifying belief \(\nu\) such that
\[
  \mathbb E_\nu[u_i]\ge0\quad(i=0,\ldots,n),
  \qquad
  \mathbb E_\nu[u_j]>0\quad\text{for some }j\in\{0,\ldots,n\}.
  \tag{NO-ACTION}
\]
Because \(u_0=R\) is deterministic, \(R>0\) satisfies the strict-gain requirement
for the auctioneer.

\paragraph{Which certifying beliefs?}
We require the certifying belief to belong to
\[
  \mathcal C:=\operatorname{co}\{P_1,\ldots,P_n\}
  =\left\{\sum_{i=1}^n\lambda_iP_i:
  \lambda_i\ge0,\ \sum_{i=1}^n\lambda_i=1\right\}.
  \tag{HULL}
\]
This is a further departure from Gilboa, Samuelson, and Schmeidler (2014,
Section 4.3), who allow any hypothetical belief. Brunnermeier, Simsek, and Xiong
(2014) use the convex hull of agents' beliefs for belief-neutral welfare comparisons.
Our certificate is existential; their efficiency and inefficiency classifications
quantify over every belief in that set. Here the choice is motivated by sequential
mechanism design: we begin with the one-box interval below and later permit a separate
product certificate at each decision, with every marginal in its corresponding range.
Certification cannot then use probabilities outside those ranges to enlarge the
revenue benchmark.

\paragraph{From individual inequalities to a revenue metric.}
The individual part of (NO-ACTION) is
\[
  \mathbb E_\nu[U_i]\ge t_i\quad(i=1,\ldots,n),
  \qquad R\ge0.
\]
Summing the agents' inequalities gives the aggregate certificate
\[
  0\le R\le\max_{\nu\in\mathcal C}
  \mathbb E_\nu\!\left[\sum_{i=1}^n U_i\right].
  \tag{NBR-CERT}
\]
An outcome is \textbf{NBR-admissible} if revenue is deterministic, every agent satisfies
weak own-belief individual rationality, and (NBR-CERT) holds. The individual
inequalities motivate this definition and are a sufficient way to verify it. The
aggregate condition is sufficient for the revenue bound needed by the benchmark:
transfers between agents cancel before evaluation, leaving aggregate allocated value.
It need not make every agent individually rational under one common belief and does
not certify an absence of betting. Strict improvement is not an additional requirement
of the aggregate definition, which also admits zero-revenue outcomes, including no action.

\paragraph{Complete ranking within the admissible class.}
For admissible outcomes \(a,b\), NBR ranks \(a\) at least as high as \(b\) exactly
when \(R(a)\ge R(b)\). Deterministic auctioneer revenue therefore gives a complete
ranking, with equal revenues receiving equal scores. The same scalar ranks outcomes
satisfying dynamic NBR below. For an equilibrium benchmark, take the supremum of
\(R\) over the specified mechanisms and their admissible equilibrium outcomes.

\subsection{One box: probability player, value player, and auctioneer}
\label{one-box-nbr}

Let \(X\in\{0,1\}\) indicate success of the single box, put
\(p_i=P_i(X=1)\), and write \(\underline p=\min_i p_i\) and
\(\overline p=\max_i p_i\). Then
\[
  \mathcal C=\{\operatorname{Ber}(\theta):
  \theta\in[\underline p,\overline p]\}.
\]
This is a single allocation decision with uncertain execution; there is no later box
whose reach must be evaluated.

Suppose the probability player \(r\) and value player \(w\) are distinct. The value
player receives value \(v>0\) upon success and then pays \(y\) to the probability
player. They pay \(t_r,t_w\) up front, and all other agents have zero payoff. Thus
\[
  U_r(X)=Xy,\qquad U_w(X)=X(v-y),\qquad u_0=R=t_r+t_w.
\]
Their own-belief individual rationality conditions are
\[
  p_r y-t_r\ge0,\qquad p_w(v-y)-t_w\ge0.
  \tag{ONE-IR}
\]
The detailed comparison with no action requires one
\(\theta\in[\underline p,\overline p]\) such that
\[
  \theta y-t_r\ge0,\qquad
  \theta(v-y)-t_w\ge0,\qquad
  R\ge0,
  \tag{ONE-CERT}
\]
with at least one inequality strict, including the possibility \(R>0\). NBR uses
(ONE-IR) and deterministic revenue together with only the summed requirement
\[
  0\le t_r+t_w\le\overline p\,v.
\]
In RWSP, \(t_w=0\). When \(0\le y\le v\),
(ONE-IR) also implies both agents' individual weak inequalities at
\(\theta=\overline p\).
For example, \(p_r=p_w=p\in(0,1)\), \(y=v\), \(t_r=pv\), and \(t_w=0\)
give both agents zero surplus under their own and certifying belief, while the
auctioneer gains \(pv>0\). Neither agent needs a strict gain.

\subsection{Dynamic NBR: The Sequential Extension of NBR}\label{the-bcr-criterion}

Sequential decisions create a reach problem absent from NBR. A high probability of
early success can justify an early payment but reduces the probability of reaching a
later allocation. The later agent may expect early failure and therefore be willing to
pay more for that allocation. A belief that justifies the early payment may consequently
fail to justify the later payment. Appendix~\ref{app:bcr} gives a two-decision example
and the full calculation.

Fix an outcome of a finite-horizon mechanism. Decision \(t\) contributes random payoff
\(U_{i,t}\) to agent \(i\), including discount and the indicator that the decision is
reached. Contingent transfers between agents balance at each decision in every state,
and agent \(i\) pays \(t_i\) up front. The dynamic NBR criterion extends NBR in two respects. First, let \(\mathcal P\)
contain the product priors whose marginal probabilities lie between the agents'
corresponding marginal beliefs:
\[
  \mathcal P=\left\{\bigotimes_k\nu_k:
  \min_iP_i(X_k=x)\le\nu_k(x)\le\max_iP_i(X_k=x)
  \text{ for every }k,x\right\}. \tag{RANGE}
\]
Second, the individual extension of the motivating certificate chooses \(\nu^t\in\mathcal P\)
for each decision \(t\) so that
\[
  \sum_t\mathbb E_{\nu^t}[U_{i,t}]-t_i\ge0
  \qquad\text{for every agent }i.
\]
Different decisions may use different priors, but within each decision one prior
evaluates every agent's payoff and the event of reaching that decision. Summing these
individual inequalities motivates the aggregate condition below.

A mechanism outcome satisfies \textbf{dynamic NBR} when three requirements hold. First,
retained revenue is deterministic and equals \(R=\sum_i t_i\ge0\). Second, every agent
has own-belief individual rationality,
\(\sum_t\mathbb E_i[U_{i,t}]\ge t_i\). Third, revenue has the
decision-by-decision certificate
\[
  R\le\sum_t\sup_{\nu\in\mathcal P}
  \mathbb E_\nu\!\left[\sum_iU_{i,t}\right]. \tag{CERT}
\]

Deterministic auctioneer revenue ranks the admissible outcomes, and own-belief
individual rationality preserves voluntary participation. The aggregate certificate is
sufficient to bound admissible revenue by certifiable allocated value; this is the
property used by the benchmark proofs. For a fixed allocation and decision structure,
adding transfers that sum to zero at each decision in every state does not increase the
right-hand side of (CERT). Such transfers can still affect own-belief participation
and therefore which outcomes qualify. The dynamic NBR definition does not require an absence of betting or
individual rationality under the certifying priors. The membership proofs establish the
stronger individual certificate, as recorded in Appendix~\ref{app:bcr}.

For one Bernoulli box, there is one decision and no later-box reach probability to
evaluate. In this case,
\[
  \mathcal P=\mathcal C
  =\{\operatorname{Ber}(\theta):
  \theta\in[\underline p,\overline p]\},
\]
so the sum over decisions in (CERT) has one term and (CERT), together with \(R\ge0\),
becomes
\[
  0\le R\le\max_{\nu\in\mathcal C}
  \mathbb E_\nu\!\left[\sum_iU_i\right],
\]
which is (NBR-CERT). Deterministic revenue and own-belief individual rationality are the
same requirements in both definitions. Thus an outcome satisfies dynamic NBR if and only if
it is NBR-admissible in the one-Bernoulli-box setting. This is the precise sense in which
dynamic NBR extends NBR. With several boxes, by contrast, \(\mathcal P\) consists of products of
the marginal beliefs in the agents' ranges; it is not in general the convex hull of the
agents' full joint priors. For outcomes with more than two possible values per coordinate,
(RANGE) retains the stated marginal bounds without asserting equality to a marginal
convex hull.

Zero-revenue outcomes are admitted by the weak aggregate definition; positive revenue
gives the auctioneer a strict gain over no action without requiring strict surplus from
any agent.

For the example in Section~\ref{example-disagreement-bet}, no value is allocated and
the transfers cancel at each decision in every outcome. Hence
\(\sum_iU_{i,t}=0\) for every decision \(t\), under every certifying prior, and the
right-hand side of (CERT) is zero. The dynamic NBR criterion therefore excludes its positive revenue
\(2\Lambda\), even though both agents satisfy own-belief individual rationality.

\subsection{Why range constraints and the aggregate certificate are needed}

Without the range restriction, a certificate could assign positive probability to
success of a box that every agent regards as certain to fail, and use its value to
justify arbitrarily large payments. Requiring that box's marginal to lie between the
agents' corresponding marginal beliefs excludes this certificate.

Finally, prohibiting up-front transfers to agents does not bound benchmark revenue.
Agents can reproduce the same bet through state-contingent transfers attached to
different decisions while all up-front payments flow to the mechanism. The resulting
revenue remains deterministic and unbounded under own-belief individual rationality.
Appendix~\ref{app:bcr} gives this construction as well. The dynamic NBR definition requires the revenue
certificate (CERT); restricting the signs of up-front transfers is insufficient.

%% file: 04-known-values.tex
\section{Revenue Guarantees for Strategic Pandora}
\label{known-values-the-revenue-guarantee}
\phantomsection\label{revenue-guarantees}

The two mechanisms face the same economic question. A high probability report can move
a box forward and increase its chance of supplying the selected action, but the same
report also determines an up-front payment. We ask whether competition forces those
payments to recover a constant fraction of the optimal revenue among mechanisms in our
admissibility class.

\phantomsection\label{strategic-pandora-form}
\phantomsection\label{the-revenue-optima}
We formalize the comparison through two benchmarks, defined pointwise in the type
profile. Let \(\mathcal A_{\mathrm K}(p,v)\) and
\(\mathcal A_{\mathrm{RV}}(p,v)\) denote the admissible mechanism--equilibrium pairs
under known and reported values, respectively. Thus \((\mathcal M,e)\) belongs to the
relevant set if \(e\) is a pure equilibrium of \(\mathcal M\) at \((p,v)\) and its
induced outcome satisfies dynamic NBR. Define
\[
  \operatorname{OPT}_{\mathrm{DNBR}}(p,v)
  :=
  \sup_{(\mathcal M,e)\in\mathcal A_{\mathrm K}(p,v)}
  \operatorname{Rev}_{\mathcal M}(e;p,v),
  \qquad
  \operatorname{OPT}_{\mathrm{RV\text{-}DNBR}}(p,v)
  :=
  \sup_{(\mathcal M,e)\in\mathcal A_{\mathrm{RV}}(p,v)}
  \operatorname{Rev}_{\mathcal M}(e;p,v).
\]

\subsection{Known box values}\label{constant-probability-competition-known}
\phantomsection\label{the-rw-revenue-lower-bound}

Here \(v_j>0\) is public and agents report only success probabilities. We isolate the
role of competition through one local condition. The profile has
\textbf{constant probability competition} \((\mathrm{PC}\text{-}K)\) if every box has
\(K\) distinct agents whose true success probabilities are at least an \(\alpha\)
fraction of the highest belief on that box. The required number of agents on each box depends on the
discount factor and is independent of the number of boxes.

The mechanism creates the same bid-shading incentive as a first-price auction: without a credible rival, the most optimistic agent can lower its report and payment while
retaining the reader role. A close probability rival creates a threshold: once revenue
falls too low relative to the box's index, that rival can report just enough to take the
role profitably. The sequential setting makes this argument substantially less local
than in a single-item auction. Taking one reader role changes the inspection order, the
probability of reaching every later box, and the payments attached to those boxes. The
proof shows that these downstream effects can still be charged to the revenue already
collected on the prefix.

Our exact dynamic NBR membership proof uses one additional selection property: every selected
reader has a nonnegative position on its box. Appendix~\ref{app:bcr} states the
equilibrium-selection rule that supplies this property, while
Appendix~\ref{app:known-pure} gives the revenue argument and exact constants.
\phantomsection\label{known-values-membership}

\phantomsection\label{thm:rw-benchmark}
\phantomsection\label{the-comparison-with-the-bcr-optimum}
\phantomsection\label{thm:known-comparison}
\textbf{Theorem 2} (Known-values comparison). \emph{For every
  \(0<\alpha\le1\) and \(0<\delta<1\), there is a finite probability depth
  \(K_0(\delta)\), independent of the number of boxes, such that at every type profile
  satisfying \((\mathrm{PC}\text{-}K_0)\), every existing pure RW equilibrium selected by
  the favorable membership tie rule satisfies}
\[
  \operatorname{Rev}_{\mathrm{RW}}
  =\Omega\!\left(
  \alpha(1-\delta)^2\operatorname{OPT}_{\mathrm{DNBR}}(p,v)
  \right).
  \tag{KV-COMP}
\]

The competition argument first gives an RW revenue lower bound for the broader
pure-equilibrium class. The favorable membership selection is used separately to show
that the induced outcome satisfies dynamic NBR.

The independence from the number of boxes is the substantive point. A longer system
creates more possible reads, more order changes, and more opportunities for strategic
reports to affect downstream reach. Nevertheless, the mechanism does not require a
growing market on every box. Discounting limits the economically relevant prefix, and a
constant local depth of probability competition is enough to control the entire
inspection sequence.

\paragraph{Why competition is necessary.}\label{why-some-competition-is-necessary}
Without a close probability rival, RW revenue can be arbitrarily small relative to the
dynamic NBR benchmark, even with two boxes. A simple voluntary direct
mechanism has a pure-equilibrium outcome satisfying dynamic NBR whose revenue approaches the
full value of the valuable box, whereas RW collects only the report needed to stay ahead
of the next belief. As the second-highest belief falls relative to the highest,
\(\operatorname{Rev}_{\mathrm{RW}}/\operatorname{OPT}_{\mathrm{DNBR}}\) converges to
zero. The familiar bid-shading obstruction of a first-price auction causes this failure through probability reports.
Appendix~\ref{app:known-separation} gives the construction and its
exact tie-selection boundary.
\phantomsection\label{prop:bcr-separation}

\paragraph{Pure-equilibrium existence and the report domain.}\label{known-existence-boundary}
The maintained probability-report domain is \([0,1]\), and the endpoint is substantive.
On the open domain \([0,1)\), a three-box instance has no pure
\(\varepsilon\)-equilibrium below a fixed positive constant, under any complete tie
rule, even though every true success probability is at most \(3/5\) (and analogous
constructions exist with arbitrarily small true probabilities). The failure is caused
by a large ratio between agents' assessments of the probability of reaching later
boxes. An agent that is more pessimistic about an early box assigns greater reach to
later boxes than the adopted reports imply. It exploits this discrepancy by
overreporting on the tail boxes. In the construction, reporting nearly one on \(B\)
drives the payment for \(C\) to zero, while the agent's true beliefs still assign
\(C\) positive reach, making the joint takeover profitable.

The closed domain removes that strict boundary deviation but does not settle universal
existence. Coordinatewise report-raising repairs increase RW revenue and admit a strict
repair potential, yet their limit need not be a Nash equilibrium: an order-changing
deviation may lose on the promoted box while gaining on a pre-existing later role. This
is why compactness and monotone repair are not enough. Appendices~\ref{app:existence-boundary}
and~\ref{app:open-domain-nonexistence} give the exact results. Universal pure-equilibrium
existence on \([0,1]\) remains open.

\phantomsection\label{known-values-mixed}
Appendix~\ref{app:known-mixed} gives the separate ordinary mixed-equilibrium revenue
bound. It is a revenue result and does not establish dynamic NBR membership for mixed equilibria.

%% file: 05-reported-values.tex
\subsection{Reported values}\label{reported-values}

\paragraph{Dynamic NBR membership under reported values.}\label{reported-pure-membership}
RWSP separates the reader and writer roles. Establishing dynamic
NBR membership therefore requires controlling the success transfer paid by a writer
distinct from the reader.

\phantomsection\label{thm:writer-only}
\textbf{Theorem 3} (Writer-only no-overbidding). \emph{At every pure equilibrium of
  RWSP in the interior probability-report domain,
  for every box \(j\) whose selected writer \(w(j)\) differs from its reader \(r(j)\),
  one has}
\[
  y_j\le v_{w(j)}^j.
  \tag{NO}
\]
Appendix~\ref{app:reported-membership} proves Theorem~3.

When the reader and writer coincide, the success transfer is internal. Otherwise,
Theorem~3 implies that the writer's net value upon success is nonnegative.

\phantomsection\label{thm:reported-membership}
\textbf{Theorem 4} (Reported-values dynamic NBR membership). \emph{Every pure equilibrium of
  RWSP in the interior probability-report domain
  induces an outcome satisfying dynamic NBR.}
Appendix~\ref{app:reported-membership} proves Theorem~4.

\paragraph{Why revenue needs two kinds of competition.}\label{reported-mixed-revenue}
\phantomsection\label{reported-constant-competition}
\phantomsection\label{reported-comparison}

Probability competition plays the same economic role on the reader side as under known
box values: a low probability payment invites a sufficiently informed rival to take the
role. It is no longer sufficient by itself. RWSP uses the
second-highest value both as the success transfer and as the value in the box's index.
If the highest-value agent has no close value rival, that second price can collapse even
when the selected write is extremely valuable.

We therefore require \textbf{qualified value competition}: on every box, the
second-highest true value is at least a \(\beta\) fraction of the highest true value. We
also require pointwise exact \textbf{writer-allocation Envy-Free}: at the realized
second price, the selected writer's true value is at least the price and every
nonwriter's true value is at most the price. Qualified value competition keeps the price
commensurate with the available value; writer-allocation Envy-Free connects the reported
second-price allocation to the agents' true values. These are different requirements,
and neither substitutes for probability competition on the reader side.

\phantomsection\label{thm:reported-constant}
\textbf{Theorem 5} (Reported-values comparison). \emph{For every
  \(0<\alpha,\beta\le1\) and \(0<\delta<1\), there is a finite probability depth
  \(K_0(\alpha,\beta,\delta)\), independent of the number of boxes, such that every
  existing pure equilibrium with interior probability reports that satisfies constant
  probability competition, qualified value competition, and pointwise exact
  writer-allocation Envy-Free satisfies}
\[
  \operatorname{Rev}_{y}
  =\Omega\!\left(
  \alpha\beta(1-\delta)^3
  \operatorname{OPT}_{\mathrm{RV\text{-}DNBR}}
  \right).
  \tag{SV-COMP}
\]

By Theorem~4, every such equilibrium outcome satisfies dynamic NBR.

This result is not a direct extension of the known-values theorem. The probability
report that moves a box forward, the value report that receives the write, and the
second value report that sets the price may come from three different agents. The proof
must therefore show simultaneously that probability competition supports the reader
deviation, qualified value competition keeps the true second value commensurate with the
highest value, and writer-allocation Envy-Free connects that true-value floor to the
realized second price. Even with this separation, the required probability depth remains
independent of the number of boxes. Appendix~\ref{app:reported-constant} gives the exact
coefficient, the explicit \(K_0\), and the sharper finite-horizon comparison.

\subsection{Scope of the Revenue Guarantees.}

The comparison guarantees in Theorems 2 and 5 apply only to equilibria in their stated
classes. Universal pure-equilibrium existence for RW on \([0,1]\) remains open, as does
existence of a
RWSP equilibrium satisfying exact writer-allocation Envy-Free.
Appendix~\ref{app:existence-boundary} discusses the RW existence boundary, and
Appendix~\ref{app:open-domain-nonexistence} proves nonexistence on the open report
domain \([0,1)\).

%% file: 07-economic-state-machine.tex
\section{Related Work}\label{related-work}

\paragraph{Pandora, reservation indices, and learning.}
Weitzman (1979) derives an optimal inspection order and stopping rule for independent
boxes with known prize distributions, inspection costs, and delay. The rule assigns a
reservation value to each box. We use Bernoulli boxes with delay and zero inspection
cost, and require search to stop at the first success. The ordering lemma gives the
allocation rule used by RW. Agents report the probabilities used to order the boxes
and determine payments.

Olszewski and Weber (2015) study when a Pandora rule is optimal for more general
objectives and relate the rule to the Gittins index for an equivalent bandit problem.
Doval (2018) allows
selection without inspection, Boodaghians et al.\ (2020) impose order constraints, and
Chawla et al.\ (2020, 2023) study correlated boxes. Esfandiari et al.\ (2019) study
online Pandora problems and their relation to prophet and bandit problems. Atsidakou
et al.\ (2022) learn contextual reservation values, including under bandit feedback.
In our model, each box is inspected at most once per round; the mechanism takes
probability reports as given and does not learn distributions from repeated
observations. Several agents may report different beliefs about the same box.

Chawla et al.\ (2026) study posted inspection prices for one sequential searcher.
Seller and buyer know the distributions, and the buyer pays for each inspection.
In RW, agents report success probabilities and pay up front according to the reported
objective term. Our benchmark maximizes deterministic revenue over mechanism--equilibrium
pairs whose outcomes satisfy dynamic NBR.

\paragraph{Dynamic mechanism design and information acquisition.}
Parkes and Singh (2003) use an MDP to implement a long-run value-maximizing policy in
Bayesian Nash equilibrium when agents arrive and depart. The dynamic pivot mechanism
implements efficient allocation under a specified stochastic model (Bergemann and
V\"alim\"aki 2010). Qiu et al.\ (2024) learn an unknown MDP over repeated episodes.
These mechanisms evaluate policies under one known or learned model. Our agents may
disagree about the same process, and we compare revenue without a common prior.

Cavallo and Parkes (2008) allocate costly deliberation in metadeliberation auctions.
In the basic model, each agent reports an MDP for deliberation about its own value.
Their extensions allow deliberation about other agents' values. Our mechanism orders
inspections of shared conditions. Every player may have a private belief and a private
value for every branch of the same process. Under reported values, the player reporting
a condition probability need not be the player that values the resulting write.

Bergemann and V\"alim\"aki (2002) and Persico (2000) study information acquisition
before a static allocation. Bergemann and V\"alim\"aki show that efficient mechanisms
provide socially efficient incentives to acquire information in their private-value
model. In Persico's model with affiliated information, a first-price auction gives
stronger incentives for continuous signal acquisition than a second-price auction.
Our model takes private beliefs and values as given. It does not model or price their
acquisition.

Porter et al.\ (2008) elicit task-success probabilities and execution costs in
fault-tolerant mechanism design. The probability concerns the agent assigned the task.
In our model, several agents hold private beliefs about the same external condition.

\paragraph{Algorithms as mechanisms and equilibrium guarantees.}
Lucier and Syrgkanis (2015) and D\"utting, Kesselheim, and Tardos (2021) obtain
equilibrium welfare guarantees by combining greedy and relax-and-round allocation
algorithms with payment rules. Hartline, Hoy, and Taggart (2014) obtain equilibrium
revenue guarantees from the prices agents face when deviating. RW combines the
discounted Weitzman order with the up-front objective-term payment rule. A probability
report changes a box's position and the probability of reaching later boxes. We bound
revenue by certifiable allocated value through dynamic NBR and compare RW revenue with the
resulting benchmark.

\paragraph{Heterogeneous beliefs and no-betting.}
Gilboa, Samuelson, and Schmeidler (2014) use a hypothetical shared belief to refine
Pareto comparisons. Their strict-gain requirement applies to every affected agent.
We include the auctioneer and allow weak gains for agents. Positive auctioneer revenue
can supply the strict gain.

For a single decision, we restrict certification to the convex hull of agents' beliefs.
Brunnermeier, Simsek, and Xiong (2014) use this set to classify belief-neutral efficiency
and inefficiency, evaluating every belief in the set. Our certificate requires one
belief in the set. NBR ranks admissible outcomes by deterministic auctioneer revenue;
the dynamic NBR criterion extends the comparison to sequential decisions
(Section~\ref{belief-compatible-revenue-and-the-prior-free-revenue-benchmark}).
The aggregate certificates bound revenue by certifiable allocated value but may admit
mechanisms containing bets. The dynamic NBR definition permits a different product prior at each decision,
with marginals in the agents' belief ranges. The same prior evaluates payoffs at that
decision and its probability of being reached.

\paragraph{Transaction sequencing and conditional execution.}
Exchanges and brokers, such as Nasdaq and on-chain order books such as Hyperliquid,
accept conditional orders in a limited language and order them by price and then time
(Harris 2003; Hyperliquid 2024). Uniswap v4 hooks provide a related form of conditional
execution, activated by a separate user calling the pool. The calling user bears the gas
cost of executing an earlier user's stored instruction (Adams et al.\ 2024).
Frequent batch auctions instead process orders at
discrete intervals (Budish, Cramton, and Shim 2015). These approaches do not
extend naturally to a composable environment where one rule may include operations
across several markets. No single price ranks a composite rule, rendering Price First
ill-defined. Submission time is also meaningless when a rule may wait for days or
weeks before its condition holds. We therefore focus on content-agnostic mechanisms,
which use only the reported condition probability and write value, not the content and
semantics of instructions.

Transaction fee mechanisms price access for an available set of transactions
(Roughgarden 2021; Chung and Shi 2023). Verifiable Sequencing Rules restrict how an
operator orders included exchange transactions and use transaction direction and
exchange-state price effects (Ferreira and Parkes 2023). We allocate condition reads
and the next transaction slot among standing instructions, using only reported
probabilities and write values.

Chitra, Ferreira, and Kulkarni (2023) use blockchains to restrict deviations by a
self-interested auctioneer. We assume execution of the stated mechanism after
admission. Event-driven contracts, active databases, keepers, and solvers already
evaluate conditions and dispatch actions (Widom and Ceri 1996; Paton and D\'iaz 1999;
Kaleem et al.\ 2021; Zhao et al.\ 2022).

%% file: acknowledgments.tex
\section*{Acknowledgments}

The first author was supported in part by two generous gifts to the Center for Research
on Computation and Society at Harvard University, both made to support research on applied
cryptography and society.

%% file: references.tex
\bookmarksetup{startatroot}
\section*{References}\label{refs}
\addcontentsline{toc}{section}{References}
Adams, Hayden, Moody Salem, Noah Zinsmeister, Sara Reynolds, Austin Adams, Will Pote,
Mark Toda, Alice Henshaw, Emily Williams, and Dan Robinson. 2024. {``Uniswap v4 Core.''}
\url{https://app.uniswap.org/whitepaper-v4.pdf}.

Aave. n.d. {``Health Factor \& Liquidations.''} Aave Help.
Accessed September 9, 2026.
\url{https://aave.com/help/borrowing/liquidations}.

Anoma. 2024. {``Anoma: An Intent-Centric Protocol for Decentralized Counterparty Discovery, Solving, and Settlement.''} \url{https://anoma.net}.

Atsidakou, Alexia, Constantine Caramanis, Evangelia Gergatsouli, Orestis
Papadigenopoulos, and Christos Tzamos. 2022. {``Contextual Pandora's Box.''}
\emph{arXiv Preprint arXiv:2205.13114}.

Axios. 2025. {``Surveying the Damage from Friday's Crypto Market Plunge.''} October 14.
\url{https://www.axios.com/2025/10/14/trump-tariffs-china-crypto}.

Bergemann, Dirk, and Juuso V\"alim\"aki. 2002. {``Information Acquisition and Efficient
Mechanism Design.''} \emph{Econometrica} 70 (3): 1007--33.
\url{https://doi.org/10.1111/1468-0262.00317}.

Bergemann, Dirk, and Juuso Välimäki. 2010. {``The Dynamic Pivot Mechanism.''} \emph{Econometrica} 78 (2): 771--89.

Better Markets. 2025. {``Request for Input on Recommendations for the {CFTC} in the
President's Working Group on Digital Asset Markets Report.''} Comment submitted to the
Commodity Futures Trading Commission, November 28.
\url{https://comments.cftc.gov/Handlers/PdfHandler.ashx?id=35776}.

Boodaghians, Shant, Federico Fusco, Philip Lazos, and Stefano Leonardi. 2020. {``Pandora's Box Problem with Order Constraints.''} In \emph{Proc. 21st ACM Conference on Economics and Computation (EC)}.

Brunnermeier, Markus K., Alp Simsek, and Wei Xiong. 2014. {``A Welfare Criterion for
Models with Distorted Beliefs.''} \emph{Quarterly Journal of Economics} 129 (4):
1753--97. \url{https://doi.org/10.1093/qje/qju025}.

Budish, Eric, Peter Cramton, and John Shim. 2015. {``The High-Frequency Trading Arms
Race: Frequent Batch Auctions as a Market Design Response.''} \emph{Quarterly Journal
of Economics} 130 (4): 1547--1621.
\url{https://doi.org/10.1093/qje/qjv027}.

Cavallo, Ruggiero, and David C. Parkes. 2008. {``Efficient Metadeliberation Auctions.''} In \emph{Proc. 23rd AAAI Conference on Artificial Intelligence}, 50--56.

Chainlink. 2023. {``Chainlink Automation.''} \url{https://chain.link/automation}.

Chawla, Shuchi, Evangelia Gergatsouli, Jeremy McMahan, and Christos Tzamos. 2023. {``Approximating Pandora's Box with Correlations.''} In \emph{Approximation, Randomization, and Combinatorial Optimization (APPROX/RANDOM)}, 275:26:1--24. LIPIcs.

Chawla, Shuchi, Dimitris Christou, Trung Dang, and Zhiyi Huang. 2026. {``Pricing
Pandora's Boxes: Revenue Maximization in Sequential Information Acquisition.''}
\emph{arXiv Preprint arXiv:2607.23359}.

Chawla, Shuchi, Evangelia Gergatsouli, Yifeng Teng, Christos Tzamos, and Ruimin Zhang. 2020. {``Pandora's Box with Correlations: Learning and Approximation.''} In \emph{Proc. 61st IEEE Symposium on Foundations of Computer Science (FOCS)}.

Chitra, Tarun, Matheus V. X. Ferreira, and Kshitij Kulkarni. 2023. {``Credible, Optimal Auctions via Blockchains.''} Cryptology ePrint Archive, Paper 2023/114. \url{https://eprint.iacr.org/2023/114}.

CFTC. 2022. {``Complaint for Injunctive and Other Equitable Relief and Civil Monetary
Penalties.''} Commodity Futures Trading Commission v. Samuel Bankman-Fried, FTX
Trading Ltd., and Alameda Research LLC, No. 1:22-cv-10503 (S.D.N.Y. December 13).
\url{https://www.cftc.gov/media/7986/enfftxtradingcomplaint121322/download}.

Chung, Hao, and Elaine Shi. 2023. {``Foundations of Transaction Fee Mechanism Design.''} In \emph{Proc. ACM-SIAM Symposium on Discrete Algorithms (SODA)}.

CoinGecko. 2025. {``What Is October 10th? {Crypto's} 10/10 Mass Market Liquidation Event.''} \url{https://www.coingecko.com/learn/october-10-crypto-crash-explained}.

DeFi Saver. 2024. {``Automation: Automated Liquidation Protection and Leverage Management.''} \url{https://defisaver.com}.

Deshpande, Amol, Lisa Hellerstein, and Devorah Kletenik. 2016. {``Approximation Algorithms for Stochastic Submodular Set Cover with Applications to Boolean Function Evaluation and Min-Knapsack.''} \emph{ACM Transactions on Algorithms} 12 (3).

Doval, Laura. 2018. {``Whether or Not to Open Pandora's Box.''} \emph{Journal of
Economic Theory} 175: 127--58.
\url{https://doi.org/10.1016/j.jet.2018.01.005}.

D\"utting, Paul, Thomas Kesselheim, and \'Eva Tardos. 2021. {``Algorithms as
Mechanisms: The Price of Anarchy of Relax-and-Round.''} \emph{Mathematics of Operations
Research} 46 (1): 317--35. \url{https://doi.org/10.1287/moor.2020.1058}.

Esfandiari, Hossein, MohammadTaghi HajiAghayi, Brendan Lucier, and Michael
Mitzenmacher. 2019. {``Online Pandora's Boxes and Bandits.''} In \emph{Proc. 33rd
AAAI Conference on Artificial Intelligence}, 1885--92.
\url{https://doi.org/10.1609/aaai.v33i01.33011885}.

Ferreira, Matheus V. X., and David C. Parkes. 2023. {``Credible Decentralized Exchange Design via Verifiable Sequencing Rules.''} In \emph{Proc. 55th Annual ACM Symposium on Theory of Computing (STOC)}. \url{https://doi.org/10.1145/3564246.3585233}.

Gelato Network. 2023. {``Gelato: A Decentralized Automation Network for Web3.''} \url{https://www.gelato.network}.

Gilboa, Itzhak, Larry Samuelson, and David Schmeidler. 2014. {``No-Betting-Pareto Dominance.''} \emph{Econometrica} 82 (4): 1405--42.
\url{https://doi.org/10.3982/ECTA11281}.

Hartline, Jason D., Darrell Hoy, and Sam Taggart. 2014. {``Price of Anarchy for Auction
Revenue.''} In \emph{Proc. 15th ACM Conference on Economics and Computation (EC)},
693--710. \url{https://doi.org/10.1145/2600057.2602878}.

Harris, Larry. 2003. \emph{Trading and Exchanges: Market Microstructure for
Practitioners}. Oxford University Press.

Hyperliquid. 2024. {``Hyperliquid: An on-Chain Order-Book Layer-1.''} \url{https://hyperliquid.xyz}.

Kaleem, Mudabbir, Keshav Kasichainula, Rabimba Karanjai, Lei Xu, Zhimin Gao, Lin Chen, and Weidong Shi. 2021. {``{EDSC}: An Event-Driven Smart Contract Platform.''} \emph{arXiv Preprint arXiv:2101.05475}.

Lucier, Brendan, and Vasilis Syrgkanis. 2015. {``Greedy Algorithms Make Efficient
Mechanisms.''} In \emph{Proc. 16th ACM Conference on Economics and Computation (EC)},
221--38.

Marshall, Michael. 2025. {``How \$3.21B Vanished in 60 Seconds: October 2025 Crypto
Crash Explained Through 7 Charts.''} Amberdata, November 3.
\url{https://blog.amberdata.io/how-3.21b-vanished-in-60-seconds-october-2025-crypto-crash-explained-through-7-charts}.

Miller, Mark S., and K. Eric Drexler. 1988. {``Markets and Computation: Agoric Open Systems.''} In \emph{The Ecology of Computation}, edited by Bernardo A. Huberman, 133--76. North-Holland.

Munagala, Kamesh, Utkarsh Srivastava, and Jennifer Widom. 2007. {``Optimization of Continuous Queries with Shared Expensive Filters.''} In \emph{Proc. 26th ACM Symposium on Principles of Database Systems (PODS)}. \url{https://doi.org/10.1145/1265530.1265561}.

Olszewski, Wojciech, and Richard Weber. 2015. {``A More General Pandora Rule?''}
\emph{Journal of Economic Theory} 160: 429--37.
\url{https://doi.org/10.1016/j.jet.2015.10.009}.

Parkes, David C., and Satinder Singh. 2003. {``An {MDP}-Based Approach to Online Mechanism Design.''} In \emph{Advances in Neural Information Processing Systems (NIPS) 16}.

Paton, Norman W., and Oscar D\'iaz. 1999. {``Active Database Systems.''}
\emph{ACM Computing Surveys} 31 (1): 63--103.
\url{https://doi.org/10.1145/311531.311623}.

PANews. 2025. {``\$19.1 Billion in Liquidation in 24 Hours, Why Did the Crypto
Market Suffer a Bloodbath?''}
\url{https://panews.io/articles/f5a1c78d-1375-41f7-875b-31cc91c61d12}.

pedma. 2025. {``What Happened on 10/10/2025?''} Trading Research Hub, November 17.
\url{https://www.tradingresearchub.com/p/what-happened-on-10102025}.

Persico, Nicola. 2000. {``Information Acquisition in Auctions.''} \emph{Econometrica}
68 (1): 135--48. \url{https://doi.org/10.1111/1468-0262.00096}.

Porter, Ryan, Amir Ronen, Yoav Shoham, and Moshe Tennenholtz. 2008. {``Fault Tolerant
Mechanism Design.''} \emph{Artificial Intelligence} 172 (15): 1783--99.
\url{https://doi.org/10.1016/j.artint.2008.06.004}.

Qiu, Shuang, Boxiang Lyu, Qinglin Meng, Zhaoran Wang, Zhuoran Yang, and Michael I.
Jordan. 2024. {``Learning Dynamic Mechanisms in Unknown Environments: A Reinforcement
Learning Approach.''} \emph{Journal of Machine Learning Research} 25 (397): 1--73.
\url{https://www.jmlr.org/papers/v25/23-0159.html}.

Reactive Network. 2024. {``Reactive Smart Contracts.''} \url{https://reactive.network}.

Roughgarden, Tim. 2021. {``Transaction Fee Mechanism Design.''} In \emph{Proc. 22nd ACM Conference on Economics and Computation (EC)}. \url{https://doi.org/10.1145/3465456.3467591}.

Sevim, Hasret Ozan, and Christof Ferreira Torres. 2026. {``Signals and Spoils:
Speculative Oracle Extractable Value in the Era of Cross-Chain Interoperability.''}
\emph{arXiv Preprint arXiv:2606.03434}.
\url{https://doi.org/10.48550/arXiv.2606.03434}.

Trillium. 2025. {``October 10, 2025 Crypto Crash: Solana Network Performance
Analysis.''}
\url{https://legacy.trillium.so/pages/oct_10_2025_crash_analysis_with_charts.html}.

Weitzman, Martin L. 1979. {``Optimal Search for the Best Alternative.''}
\emph{Econometrica} 47 (3): 641--54.
\url{https://doi.org/10.2307/1910412}.

Widom, Jennifer, and Stefano Ceri, eds. 1996. \emph{Active Database Systems: Triggers
and Rules for Advanced Database Processing}. Morgan Kaufmann.

Zhao, Zihan, Sidi Mohamed Beillahi, Ryan Song, Yuxi Cai, Andreas Veneris, and Fan Long. 2022. {``{SigVM}: Enabling Event-Driven Execution for Truly Decentralized Smart Contracts.''} In \emph{Proc. ACM on Programming Languages (OOPSLA)}. \url{https://doi.org/10.1145/3563312}.

%% file: appendices/appendix-a-discounted-pandora-ordering.tex
\appendixsection{Discounted Pandora Ordering}{app:discounted-pandora}

\textbf{Lemma 1 (Discounted Weitzman optimality).}
\emph{Among policies that execute the first successful box and then stop,
      ordering the boxes in weakly decreasing order of \(z_j\) maximizes \(W_\delta\).}

\emph{Proof.}
Consider two adjacent boxes \(a,b\), followed by a
continuation of conditional value \(C\). The contribution of the continuation is the
same under the orders \(a,b\) and \(b,a\), because in both cases it is multiplied by
\(\delta^2(1-p_a)(1-p_b)\). Subtracting the remaining terms gives
\[
  \begin{aligned}
    W_{ab}-W_{ba}
     & =
    p_av_a(d+\delta p_b)-p_bv_b(d+\delta p_a) \\
     & =(d+\delta p_a)(d+\delta p_b)
    \left(
    \frac{p_av_a}{d+\delta p_a}
    -
    \frac{p_bv_b}{d+\delta p_b}
    \right),
  \end{aligned}
\]
where \(d:=1-\delta\). Thus \(a\) weakly precedes \(b\) exactly when
\(z(p_a,v_a)\ge z(p_b,v_b)\). Repeated adjacent swaps prove that decreasing index order
maximizes \(W_\delta\). Since the Bernoulli model awards the execution value of the
first successful box and later inspection only delays that award, search terminates at
the first success. This proves the lemma.

%% file: appendices/appendix-b-restricted-nbir.tex
\appendixsection{Dynamic No-Betting Revenue: Certificates, Caps, and Known-Values Membership}{app:bcr}

\subsection{Tie rules and equilibrium selection}\label{app:tie-rules}

An \textbf{admissible report-blind tie rule} is deterministic. It may depend on true
types, public values, agent and box identities, tied-agent sets, and selected
agent--box combinations, but not on the numerical level shared inside a tie. The
\textbf{favorable membership tie rule} is the narrower selection among admissible rules
for which the boxwise deviation argument rules out a negative selected probability
position at every selected pure equilibrium. This is the selection used in the
known-values dynamic NBR membership statement. For an implementation, fixed priorities and tie
rules must depend only on recorded or public inputs; the broader type-dependent class is
used only for analytical equilibrium selection.

\subsection{Exact Strategic Pandora certificates}

The decision-by-decision individual inequalities extend the motivation from the
no-action comparison including the auctioneer in Section~\ref{the-nbr-criterion}.
Both membership proofs establish this stronger route to the aggregate dynamic NBR requirement
by choosing \(\nu^t\in\mathcal P\) for each decision \(t\) so that
\(\sum_t\mathbb E_{\nu^t}[U_{i,t}]-t_i\ge0\) for every agent \(i\). Summing over
agents implies (CERT). The benchmark needs only (CERT), which bounds revenue by
certifiable aggregate allocation. For \(R>0\), the auctioneer already has a strict
gain over no action, without requiring strict certified surplus from any agent;
\(R=0\) is admitted by the weak aggregate definition.

Fix an outcome order \(\pi\). Put
\[
  \underline p_k:=\min_i p_i^k,
  \qquad
  \overline p_k:=\max_i p_i^k.
\]
Every box-specific product prior \(\nu^j\) whose marginal probabilities
\(\theta_k^j\) lie in \([\underline p_k,\overline p_k]\) satisfies
\[
  L_j\le J_j(\nu^j)\le U_j,
  \tag{SA4}
\]
where
\[
  L_j:=
  \delta^{t(j)-1}\prod_{k\prec_\pi j}(1-\overline p_k)\underline p_j,
  \qquad
  U_j:=
  \delta^{t(j)-1}\prod_{k\prec_\pi j}(1-\underline p_k)\overline p_j.
\]
Indeed, each preceding failure factor lies between
\(1-\overline p_k\) and \(1-\underline p_k\), and the target success factor lies
between \(\underline p_j\) and \(\overline p_j\). Multiplying the nonnegative
inequalities gives both exact edges.

\subsection{Why the Dynamic NBR Restrictions Are Needed}

The examples below use boundary beliefs to show why dynamic NBR restricts certifying beliefs to the marginal ranges, permits a different certifying prior at each decision, and imposes the aggregate revenue certificate. The examples technically lie outside the interior true-belief domain that is used in the main theorems, but the boundary beliefs make the role of each restriction clear.

\subsubsection{Unrestricted certifying beliefs}
Without the range restriction, a certificate can rely on an event that every agent
regards as impossible. Consider two agents, two boxes ordered \(A,B\), a discount factor
\(0<\delta<1\), and \(\Lambda>1\). Let
\[
  v_A=\Lambda+1,
  \qquad v_B=1,
\]
and let the agents' beliefs be
\[
  (p_1^A,p_1^B)=(0,1),
  \qquad
  (p_2^A,p_2^B)=(0,0).
\]
Allocate the value of each successful box to agent \(1\), charge
\(t_1=\Lambda\) and \(t_2=0\), and set
\[
  \beta_{1B}=-\frac{\Lambda}{\delta},
  \qquad
  \beta_{2B}=\frac{\Lambda}{\delta},
  \qquad
  \beta_{1A}=\beta_{2A}=0.
\]
If \(B\) is reached and succeeds, agent \(2\) pays \(\Lambda/\delta\) to agent
\(1\). Agent \(1\) expects the transfer, whereas agent \(2\) believes it will never
occur. Their own-belief utilities are
\[
  U_1^{\mathrm{own}}=\delta-\Lambda+\Lambda=\delta>0,
  \qquad
  U_2^{\mathrm{own}}=0,
\]
while deterministic revenue is \(R=\Lambda\).

If the range condition is removed, a certifying prior can assign probability one to
success of \(A\), although both agents assign that event probability zero. The value
\(\Lambda+1\) then certifies the up-front charge, and success of \(A\) prevents the
contingent transfer on \(B\) from being reached. Certification and own-belief
individual rationality hold for every \(\Lambda\), so revenue is unbounded. The range
restriction excludes exactly this fictitious source of value.

\subsubsection{One prior for the complete outcome}
A single range-constrained product prior for the entire sequential outcome is too
restrictive. Let \(\delta=0.9\), \(v_A=v_B=1\), and consider four agents with beliefs
\[
  p_1=p_2=(0.5,0),
  \qquad
  p_3=p_4=(0,0.1).
\]
The first pair reports \((0.5,0)\), and the second reports \((0,0.2)\). RW inspects
\(A\) before \(B\). The selected reader on \(A\) receives expected value \(0.5\) and
pays \(0.5\). The reader on \(B\) believes that \(A\) certainly fails, so its expected
discounted value and its up-front payment are both
\[
  0.9(1-0)(0.1)=0.9(1-0.5)(0.2)=0.09.
\]
The report profile is a pure equilibrium and RW revenue is \(0.59\).

Any single product prior inside the agents' belief ranges has
\(0\le\theta_A\le0.5\) and \(0\le\theta_B\le0.1\). Its expected allocated value is at
most
\[
  \theta_A+0.9(1-\theta_A)\theta_B
  \le0.5+0.9(1-0.5)(0.1)
  =0.545<0.59.
\]
No single range-constrained prior can therefore certify the outcome. Decision-specific
priors can: use \((\theta_A^A,\theta_B^A)=(0.5,0)\) for decision \(A\) and
\((\theta_A^B,\theta_B^B)=(0,0.1)\) for decision \(B\). Their certified masses are
\(0.5\) and \(0.09\), exactly the mechanism revenue. The obstruction is reach. The
reader on \(B\) is pessimistic about \(A\) and therefore assigns more probability to
reaching \(B\) than a single coordinatewise optimistic prior does.

\subsubsection{Prohibiting up-front transfers}
Requiring all up-front payments to flow to the mechanism also fails to bound benchmark
revenue. Let both boxes have value one, order them as \(A,B\), and take two agents with
beliefs
\[
  (p_1^A,p_1^B)=(1,0),
  \qquad
  (p_2^A,p_2^B)=(0,1).
\]
Allocate no box value and charge \(t_1=t_2=\Lambda\). Set
\[
  (\beta_{1A},\beta_{2A})=(-\Lambda,\Lambda),
  \qquad
  (\beta_{1B},\beta_{2B})=
  \left(\frac{\Lambda}{\delta},-\frac{\Lambda}{\delta}\right).
\]
The contingent payments cancel at each box, so retained revenue is the deterministic
amount \(2\Lambda\). Under agent \(1\)'s belief, \(A\) succeeds and its contingent
receipt offsets its up-front payment. Under agent \(2\)'s belief, \(A\) fails and the
discounted receipt on \(B\) offsets its up-front payment. Both agents obtain zero
own-belief utility, while revenue grows without bound. Because no value is allocated,
the aggregate bound is zero and the outcome cannot enter the dynamic NBR benchmark.
Thus a sign restriction on up-front payments is not sufficient to bound benchmark
revenue. The aggregate certificate supplies that restriction without requiring an
absence of betting in admitted mechanisms.

\subsection{Revenue caps}

The cap uses only aggregate certification. Contingent transfers between agents cancel
decision by decision in every state, leaving aggregate allocation.

Put \(p_j^{[1]}:=\max_i p_i^j\) and
\(TW:=W_\delta(p^{[1]},v)\). For reported values, put
\(p_j^{\max}:=\max_i p_i^j\), \(v_j^{\max}:=\max_i v_i^j\), and
\(W^{\max}:=W_\delta(p^{\max},v^{\max})\). Finally, put
\[
  G_M(\delta):=\sum_{t=1}^M\delta^{t-1}\le\frac{1}{1-\delta}.
\]
Suppose first that values are public. The aggregate certificate and statewise
feasibility imply the following, where \(V_{ij}\) is agent \(i\)'s discounted
allocation contribution from box \(j\):
\[
  R
  \le
  \sum_j\sup_{\nu\in\mathcal P}
  \mathbb E_\nu\!\left[\sum_iV_{ij}\right]
  \le
  \sum_jU_jv_j.
\]
If box \(j=\pi_t\), its upper edge is at most
\(\delta^{t-1}p_j^{[1]}\). Hence
\[
  R
  \le
  \sum_{t=1}^M\delta^{t-1}p_{\pi_t}^{[1]}v_{\pi_t}
  \le
  \sum_{t=1}^M\delta^{t-1}TW
  =
  G_M(\delta)TW.
\]
Here the second inequality follows because placing any box \(j\) first in the
true-Weitzman objective gives \(p_j^{[1]}v_j\le TW\).
Taking the appropriate supremum over pure equilibria whose outcomes satisfy dynamic NBR and over
mechanisms proves
\[
  \operatorname{OPT}_{\mathrm{DNBR}}\le G_M(\delta)TW.
  \tag{B.1}
\]

Under reported values, statewise aggregate allocation on success of box \(j\)
is at most \(v_j^{\max}\). Repeating the proof gives
\[
  \operatorname{OPT}_{\mathrm{RV\text{-}DNBR}}
  \le
  G_M(\delta)W^{\max}.
  \tag{B.2}
\]

\subsection{Exact membership under known box values}

Fix an existing pure RW equilibrium selected by the favorable membership tie rule. Let
\[
  m_j:=D_j(q)\widehat q_j
\]
be the mass priced on box \(j\). The favorable rule gives the selected reader a
nonnegative box position:
\[
  D_{r(j),j}p_{r(j)}^j-m_j\ge0.
  \tag{B.3}
\]
The reader's true mass is at most \(U_j\), so (B.3) gives \(m_j\le U_j\).

For each box \(j\), choose the product prior
\[
  \theta_k^j=
  \begin{cases}
    \overline p_j,  & k=j,    \\
    \underline p_k, & k\ne j.
  \end{cases}
  \tag{B.4}
\]
Every marginal lies in its coordinatewise range, and
\(J_j(\nu^j)=U_j\). Since \(L_j\le U_j\), this certificate satisfies both exact edges
of (SA4), while \(J_j(\nu^j)\ge m_j\).

Agent \(i\)'s up-front payment is
\[
  t_i=\sum_{j:r(j)=i}m_jv_j.
\]
Under the box-specific certificates,
\[
  \sum_j\mathbb E_{\nu^j}[V_{ij}]-t_i
  =
  \sum_{j:r(j)=i}(U_j-m_j)v_j
  \ge0.
  \tag{B.5}
\]
Thus the stronger agent-by-agent certificate holds. Equilibrium opt-out gives
own-belief individual rationality. RW has no terminal inter-agent transfers, its
revenue is deterministic, and its aggregate allocation equals the produced value
statewise. Equations (B.3)--(B.5) prove the known-values membership claim in
Section~\ref{constant-probability-competition-known}.

%% file: appendices/appendix-c-pure-rw-revenue.tex
\appendixsection{Pure RW Revenue under Constant Probability Competition}{app:known-pure}

\textbf{Theorem 2} (Known-values comparison). \emph{For every
  \(0<\alpha\le1\) and \(0<\delta<1\), there is a finite probability depth
  \(K_0(\delta)\), independent of the number of boxes, such that at every type profile
  satisfying \((\mathrm{PC}\text{-}K_0)\), every existing pure RW equilibrium selected by
  the favorable membership tie rule satisfies}
\[
  \operatorname{Rev}_{\mathrm{RW}}
  =\Omega\!\left(
  \alpha(1-\delta)^2\operatorname{OPT}_{\mathrm{DNBR}}(p,v)
  \right).
  \tag{KV-COMP}
\]

\emph{Proof.}
Fix an existing pure RW equilibrium with probability reports in \([0,1]\) and an
admissible report-blind tie rule as defined in
Appendix~\ref{app:tie-rules}. Under
\((\mathrm{PC}\text{-}K_0)\), the intermediate revenue inequalities below apply to
this equilibrium. For equilibria selected by the favorable membership tie rule, the
final inequality gives the comparison with dynamic NBR.

Put \(d:=1-\delta\), \(p_j^{[1]}:=\max_i p_i^j\),
\(TW:=W_\delta(p^{[1]},v)\), and define
\[
  Z_i:=\max_j z(p_i^j,v_j),
  \qquad
  Z^*:=\max_jz(p_j^{[1]},v_j).
\]
Let \(Z^{[r]}\) be the \(r\)-th largest \(Z_i\), with unavailable ranks set to zero.
For \(1\le L\le M\), put
\[
  \tau_L:=
  \begin{cases}
    \delta^L, & L<M, \\
    0,        & L=M.
  \end{cases}
\]

\subsection{Two bounds}

For every fixed order \(\pi\),
\[
  W_\delta(p_i,v;\pi)\le Z_i,
  \qquad
  TW\le Z^*,
  \qquad
  \max_iZ_i=Z^*.
  \tag{C.1}
\]
For the first inequality, work backward through \(\pi\). If a suffix has value at most
\(Z_i\), adding a box of probability \(p\) and value \(v\) gives at most
\[
  pv+\delta(1-p)Z_i
  \le(d+\delta p)Z_i+\delta(1-p)Z_i
  =Z_i.
\]
The second inequality is identical with the boxwise maximum probabilities, and the
third follows by interchanging the two maxima.

\subsection{The position inequality}

\textbf{Lemma C.1.} For every \(0\le L\le M\), with \(\tau_0:=1\),
\[
  \operatorname{Rev}_{\mathrm{RW}}
  \ge(d-\tau_L)_+Z^{[L+1]}.
  \tag{C.2}
\]

\emph{Proof.}
At most \(L\) distinct agents own reader roles in the first \(L\) equilibrium
positions. Among the \(L+1\) agents with the largest \(Z_i\)'s, choose \(h\) who owns
none of these roles, so \(Z_h\ge Z^{[L+1]}\). Let box \(B\) attain \(Z_h\), and let
\(\widehat Z\) be the leading reported index. The first payment gives
\[
  \operatorname{Rev}_{\mathrm{RW}}\ge d\widehat Z.
  \tag{C.3}
\]
When \(\widehat Z\ge Z_h\), this proves (C.2). Otherwise, all reader roles of \(h\)
lie after position \(L\). Removing other agents' roles and closing the resulting gaps,
then applying (C.1), gives
\[
  U_h\le\tau_LZ_h.
  \tag{C.4}
\]
Agent \(h\) sets every non-target report to zero and reports on \(B\) just above the
level \(q\) satisfying \(z(q,v_B)=\widehat Z\). Since
\(z(p_h^B,v_B)=Z_h>\widehat Z\), we have \(q<p_h^B\), and the target gain approaches
\[
  (p_h^B-q)v_B
  \ge d(Z_h-\widehat Z).
  \tag{C.5}
\]
Every role retained at report zero is nonnegative. The Nash inequality, (C.4), and
(C.5) give
\[
  d(Z_h-\widehat Z)\le\tau_LZ_h.
\]
Adding (C.3) proves (C.2). For \(L=M\), agent \(h\) owns no role and the same argument
uses \(\tau_M=0\). \(\square\)

\subsection{Using constant probability competition}

Let \(B^*\) attain \(Z^*\), and choose the \(K\) agents supplied by
\((\mathrm{PC}\text{-}K)\) on \(B^*\). Since
\[
  z(\alpha p,v)\ge\alpha z(p,v),
\]
each of these agents has \(Z_i\ge\alpha Z^*\). Thus, for
\(1\le L\le\min\{M,K-1\}\),
\[
  Z^{[L+1]}\ge\alpha Z^*.
  \tag{C.6}
\]
Equations (C.1), (C.2), and (C.6) imply
\[
  \operatorname{Rev}_{\mathrm{RW}}
  \ge
  Q_{M,K}^{\mathrm{PC}}(\alpha,\delta)TW,
\]
where
\[
  Q_{M,K}^{\mathrm{PC}}(\alpha,\delta)
  :=
  \alpha\max_{1\le L\le\min\{M,K-1\}}(d-\tau_L)_+.
  \tag{C.7}
\]

For the finite-\(M\) companion bound, define
\[
  C_1:=\frac1\alpha,
  \qquad
  C_t:=
  \max\left\{
  C_{t-1}+\frac{2}{\alpha d},
  \left(1+\frac2\alpha\right)C_{t-1}
  \right\}.
  \tag{C.8}
\]
If the equilibrium order is \((j_1,\ldots,j_M)\), put
\[
  E_t:=p_{j_t}^{[1]}v_{j_t},
  \qquad
  A_t:=\sum_{s=1}^tE_s.
\]
A current nonreader on \(j_t\) has probability at least
\(\alpha p_{j_t}^{[1]}\). If the report needed to move \(j_t\) first is infeasible or
at least half this probability, the first payment bounds \(E_t\) by
\(2\operatorname{Rev}_{\mathrm{RW}}/(\alpha d)\). Otherwise the deviation creates
target utility at least \(\alpha E_t/2\), so the preceding prefix must have value at
least that amount. Induction using the two cases in (C.8) gives
\[
  A_t\le C_t\operatorname{Rev}_{\mathrm{RW}}.
  \tag{C.9}
\]
Consequently
\[
  \operatorname{Rev}_{\mathrm{RW}}
  \ge
  \widetilde c_{M,K}^{\mathrm{PC}}(\alpha,\delta)TW,
  \qquad
  \widetilde c_{M,K}^{\mathrm{PC}}
  :=
  \max\left\{\frac1{C_M},Q_{M,K}^{\mathrm{PC}}\right\}.
  \tag{C.10}
\]

\subsection{Constant depth and the restricted-class comparison}

Define
\[
  L_0:=\min\{\ell\ge1:\delta^\ell\le d/2\},
  \qquad
  K_0:=\max\{3,L_0+1\}.
  \tag{C.11}
\]
Choose \(L=\min\{M,L_0\}\) in (C.7). Then \(L+1\le K_0\), and either \(L=M\),
so \(\tau_L=0\), or \(\tau_L\le d/2\). Therefore every pure equilibrium under
\((\mathrm{PC}\text{-}K_0)\) satisfies
\[
  \operatorname{Rev}_{\mathrm{RW}}
  \ge\frac{\alpha(1-\delta)}2TW.
  \tag{C.12}
\]
Combining (C.12) with (B.1) proves
\[
  \operatorname{Rev}_{\mathrm{RW}}
  \ge
  \frac{\alpha(1-\delta)}{2G_M(\delta)}
  \operatorname{OPT}_{\mathrm{DNBR}}
  \ge
  \frac{\alpha(1-\delta)^2}{2}
  \operatorname{OPT}_{\mathrm{DNBR}}.
  \tag{C.13}
\]
Equation (C.12) is the exact intermediate revenue bound. For equilibria selected by the
favorable membership tie rule, (C.13) is the exact form of Theorem~2.

%% file: appendices/appendix-d-ordinary-mixed-rw-revenue.tex
\appendixsection{Ordinary Mixed RW Revenue}{app:known-mixed}

We show that ordinary mixed RW equilibria under known box values satisfy a
configuration-dependent lower bound on expected revenue. Under constant probability
competition with a number of agents depending only on \(\delta\), expected revenue is
an \(\Omega(\alpha(1-\delta)^2)\) fraction of the nonstrategic benchmark.

Fix an ordinary mixed RW equilibrium
\(\sigma=\prod_i\sigma_i\) under known box values. Put
\(d:=1-\delta\), \(p_j^{[1]}:=\max_i p_i^j\),
\(TW:=W_\delta(p^{[1]},v)\), \(Z_i:=\max_jz(p_i^j,v_j)\), and
\(Z^*:=\max_jz(p_j^{[1]},v_j)\). Let \(Z^{[r]}\) be the \(r\)-th largest
\(Z_i\), and put \(\tau_L:=\delta^L\) for \(L<M\) and \(\tau_M:=0\).
Let
\[
  R:=\mathbb E_\sigma[\operatorname{Rev}_{\mathrm{RW}}],
\]
let \(u_i\) be agent \(i\)'s equilibrium expected utility, and put
\(c_0:=1-e^{-1}\). The all-zero report gives \(u_i\ge0\).

Fix an agent \(i\) and target box \(B\). Let
\[
  Z:=z(p_i^B,v_B),
  \qquad
  T_i(q_{-i})
  :=
  \max_jz\!\left(\max_{h\ne i}q_h^j,v_j\right).
\]
For every realized profile, the first payment gives
\(\operatorname{Rev}_{\mathrm{RW}}\ge dT_i\), hence
\[
  R\ge d\,\mathbb E[T_i].
  \tag{D.1}
\]
For a fixed \(0<x<Z\), let \(q(x)\) satisfy \(z(q(x),v_B)=x\). The fixed deviation
reports \(q(x)\) on \(B\) and zero elsewhere. Since \(x<Z\), strict monotonicity
of \(z(\cdot,v_B)\) gives \(q(x)<p_i^B\). On \(\{T_i<x\}\), box \(B\) is first
and agent \(i\) is its reader. The deviation utility on this event is at least
\[
  (p_i^B-q(x))v_B
  \ge d(Z-x).
\]
On \(\{T_i\ge x\}\), the deviation utility is nonnegative if \(B\) is first. If
\(B\) is later, its payment is at most
\(\delta q(x)v_B\le\delta x\le\delta T_i\). Hence the deviation utility is bounded
below by
\[
  d(Z-x)\mathbf 1_{\{T_i<x\}}-\delta T_i.
\]
The Nash inequality and (D.1) give
\[
  u_i+\frac{\delta}{d}R
  \ge
  d(Z-x)\Pr(T_i<x).
  \tag{D.2}
\]
Put \(C:=R/d\) and
\[
  a:=\frac{u_i+\delta C}{dZ}.
\]
If \(a\ge1\), then \((u_i+C)/(dZ)\ge1\). Suppose \(0<a<1\). Equation (D.2)
gives
\[
  \Pr(T_i<x)\le\frac{aZ}{Z-x}\qquad(0<x<Z).
\]
Using (D.1) and integrating the corresponding lower bound on
\(\Pr(T_i\ge x)\) over \(0\le x\le(1-a)Z\) gives
\[
  \frac CZ
  \ge
  1-a-a\log(1/a).
\]
If \(a=0\), equation (D.2) gives \(\Pr(T_i<x)=0\) for every \(0<x<Z\), so
\(C/Z\ge1\) by (D.1). For \(0<a<1\),
\[
  \frac{u_i+C}{dZ}=a+\frac CZ
\]
and \(a\log(1/a)\le e^{-1}\). The cases \(a=0\) and \(a\ge1\) give the
same lower bound directly. Therefore
\[
  u_i+\frac Rd\ge c_0d\,z(p_i^B,v_B).
  \tag{D.3}
\]

Choose for each agent a box attaining \(Z_i=\max_jz(p_i^j,v_j)\). Applying (D.3)
to this box gives
\[
  u_i+\frac Rd\ge c_0dZ_i.
\]
Fix \(1\le K\le n\) and \(1\le L\le M\), and let \(H_K\) contain the \(K\)
agents with the largest \(Z_i\)'s. For every realized report profile with induced
order \(\pi\), agent \(i\)'s gross payoff is at most
\(W_\delta(p_i,v;\pi)\le Z_i\), by the first bound in
Appendix~\ref{app:known-pure}. If agent \(i\) owns no reader role in the first
\(L\) positions, its gross payoff is at most
\(\tau_LZ_i\). At most \(L\) agents in \(H_K\) own a reader role in these
positions. Up-front payments are nonnegative. Summing the resulting pointwise bounds
and taking expectations gives
\[
  \sum_{i\in H_K}\frac{u_i}{Z_i}
  \le
  K\tau_L+(1-\tau_L)L.
  \tag{D.4}
\]
Divide (D.3) by \(Z_i\), sum over \(H_K\), and use \(Z_i\ge Z^{[K]}\). This proves
\[
  R
  \ge
  d\left(
  c_0d-\tau_L-(1-\tau_L)\frac LK
  \right)_+Z^{[K]}.
  \tag{D.5}
\]
Equation (D.5) is the configuration-dependent ordinary mixed bound.

Assume the constant probability competition condition
\((\mathrm{PC}\text{-}K)\) from
Section~\ref{constant-probability-competition-known}. Let \(B^*\) attain \(Z^*\), and
choose the \(K\) agents supplied by the condition on \(B^*\). Since
\[
  z(\alpha p,v)\ge\alpha z(p,v),
\]
each of these agents has \(Z_i\ge\alpha Z^*\). Hence
\(Z^{[K]}\ge\alpha Z^*\ge\alpha TW\), and (D.5) gives, for every
\(1\le L\le M\),
\[
  R
  \ge
  \alpha d
  \left(
  c_0d-\tau_L-(1-\tau_L)\frac LK
  \right)_+TW.
  \tag{D.6}
\]

Define
\[
  L_0^{\mathrm{mix}}
  :=
  \min\left\{\ell\ge1:\delta^\ell\le\frac{c_0d}{4}\right\},
  \qquad
  K_0^{\mathrm{mix}}
  :=
  \left\lceil\frac{4L_0^{\mathrm{mix}}}{c_0d}\right\rceil.
  \tag{D.7}
\]
Under \((\mathrm{PC}\text{-}K_0^{\mathrm{mix}})\), choose
\(L=\min\{M,L_0^{\mathrm{mix}}\}\). Then
\(\tau_L\le c_0d/4\) and \(L/K_0^{\mathrm{mix}}\le c_0d/4\), so (D.6) gives
\[
  R
  \ge
  \frac{c_0}{2}\alpha d^2TW
  =\Omega\!\left(\alpha(1-\delta)^2TW\right).
  \tag{D.8}
\]
Equation (D.8) establishes the expected-revenue bound for ordinary mixed equilibria.
The paper's dynamic NBR membership results apply to pure equilibria.

%% file: appendices/appendix-e-two-box-separation.tex
\appendixsection{A Two-Box Separation}{app:known-separation}

We show that RW revenue can become arbitrarily small relative to the dynamic NBR
benchmark in a two-box instance when one box lacks a close probability rival.

Fix \(0<\delta<1\), put \(d:=1-\delta\), and let \(r>0\) tend to zero. There
are two boxes \(A,B\) and two agents with
\[
  p_1=(1,1),
  \qquad
  p_2=(r,1),
  \qquad
  v_B=1.
\]
Set
\[
  \kappa:=d+\delta r,
  \qquad
  v_A:=\frac{\kappa^2}{r(d+\delta\kappa)}.
  \tag{E.1}
\]
At the report profile
\[
  q_1=(r,0),
  \qquad
  q_2=(r,0),
  \tag{E.2}
\]
select agent \(1\) on \(A\), agent \(2\) on \(B\), and inspect \(A\) first.

Agent \(1\)'s utility is \((1-r)v_A\). Lowering its report on \(A\) loses this role,
raising it increases its payment, and a positive report on \(B\) that leaves \(B\)
second has zero subjective reach. Moving \(B\) first requires a report at least
\(\kappa\), because
\[
  \frac{rv_A}{d+\delta r}
  =
  \frac{\kappa}{d+\delta\kappa}.
\]
Such a deviation yields utility at most one, whereas
\((1-r)v_A>1\) for all sufficiently small \(r\).

Agent \(2\)'s utility is \(\delta(1-r)\). If \(B\) remains second and agent \(2\)
does not win \(A\), a positive report \(b\) on \(B\) gives at most
\(\delta(1-r)(1-b)\). If it also wins \(A\), its report there must satisfy
\(a>r\), and its total utility is at most
\[
  \delta(1-r)+(r-a)v_A-\delta(1-a)b
  \le\delta(1-r).
\]
Moving \(B\) first requires a report at least \(\kappa\), and gives utility at most
\[
  1-\kappa=\delta(1-r).
\]
If agent \(2\) also wins the later \(A\), its subjective reach there is zero because it
believes \(B\) succeeds surely, while its payment is nonnegative. The analogous
later-box reach is zero for agent \(1\). Thus these cases also cover simultaneous
two-coordinate deviations, and (E.2) is a pure equilibrium for every sufficiently
small \(r\).

RW revenue at this equilibrium is \(rv_A\). For any \(\lambda<1\), consider the
voluntary direct mechanism that inspects only \(A\), allocates its value to agent \(1\),
and charges \(\lambda v_A\) if agent \(1\) opts in. The agent obtains positive utility.
For its box-\(A\) certificate, take the product prior whose \(A\)-marginal is \(1\);
choose every unused marginal within its coordinatewise range. Its mass lies between the
exact edges \(r\) and \(1\), and the agent-by-agent certificate utility is
\((1-\lambda)v_A\ge0\). Revenue is deterministic and allocation is feasible. The
outcome therefore has exact dynamic NBR membership. Taking the supremum as
\(\lambda\uparrow1\) gives
\[
  \operatorname{OPT}_{\mathrm{DNBR}}\ge v_A.
\]
Consequently
\[
  \frac{\operatorname{Rev}_{\mathrm{RW}}}
  {\operatorname{OPT}_{\mathrm{DNBR}}}
  \le r
  =
  \frac{p_A^{[2]}}{p_A^{[1]}}
  \longrightarrow0.
\]
This proves the two-box separation discussed in
Section~\ref{why-some-competition-is-necessary}.

%% file: appendices/appendix-f-reported-values-membership.tex
\appendixsection{Writer-Only No-Overbidding and Reported-Values Membership}{app:reported-membership}

\textbf{Theorem 3} (Writer-only no-overbidding). \emph{At every pure equilibrium of
  RWSP in the interior probability-report domain,
  for every box \(j\) whose selected writer \(w(j)\) differs from its reader \(r(j)\),
  one has}
\[
  y_j\le v_{w(j)}^j.
  \tag{NO}
\]

\emph{Proof.}
Fix a pure equilibrium in the interior probability-report domain of Theorem~3, a box
\(A\), and a selected writer \(w\ne r(A)\). Write
\[
  y:=y_A,
  \qquad
  \widehat q:=\widehat q_A,
  \qquad
  H:=\{i\ne w:b_i^A\ge y\}.
\]
Suppose for contradiction that \(y>v_w^A\). The writer contribution of \(A\) to
\(w\)'s utility is strictly negative:
\[
  S_{w,A}(p_w;\pi)p_w^A(v_w^A-y)<0.
  \tag{F.1}
\]
We show that \(w\) can surrender this role while preserving the second price, adopted
probability, complete index vector, and order.

If \(|H|\ge2\), lower \(b_w^A\) below \(y\). Two other reports preserve the second
price, all probability reports remain fixed, and the reader remains fixed. The negative
term (F.1) disappears.

It remains to consider \(H=\{s\}\). Then
\[
  b_s^A=y,
  \qquad
  b_k^A<y\quad(k\notin\{w,s\}).
  \tag{F.2}
\]
If \(q_w^A<q_s^A\), setting \(b_w^A=y\) makes \(s\) the writer while preserving all
other relevant quantities. If \(q_w^A=q_s^A<\widehat q\), additionally lower
\(q_w^A\) by an arbitrarily small amount; another agent preserves the adopted
probability, while the effect on later personalized payments vanishes with the
perturbation.

Suppose \(q_s^A<q_w^A<\widehat q\). While \(q_w^A\) stays below
\(\widehat q\), define the nonnegative coefficient
\[
  K_w^y(A;\pi)
  :=
  \sum_{\substack{j:A\prec_\pi j\\r(j)=w}}
  \delta^{t_\pi(j)-1}
  \prod_{\substack{k\prec_\pi j\\k\ne A}}(1-q_w^k)
  q_w^j y_j.
  \tag{F.3}
\]
If (F.3) is positive, a small increase in \(q_w^A\) is already profitable. If it is
zero, set \(b_w^A=y\) and \(q_w^A=0\). Agent \(s\) becomes the writer, another agent
preserves \(\widehat q\), and (F.3) shows that the finite probability change has no
later-payment cost.

Finally, if \(q_w^A=\widehat q\), the reader rule and \(w\ne r(A)\) imply
\[
  r(A)=s,\qquad b_w^A=b_s^A=y,\qquad q_w^A=q_s^A=\widehat q.
\]
Lower \(q_w^A\) slightly. Agent \(s\) remains reader and becomes writer; all indices
and the order remain fixed, and the later-payment effect vanishes with the
perturbation. Every case removes the fixed negative term (F.1) at arbitrarily small or
zero other cost, contradicting equilibrium. This proves Theorem~3.

\textbf{Theorem 4} (Reported-values dynamic NBR membership). \emph{Every pure equilibrium of
  RWSP in the interior probability-report domain
  induces an outcome satisfying dynamic NBR.}

\emph{Proof.}
Fix the same pure equilibrium, and put \(v_j^{\max}:=\max_i v_i^j\). Let
\[
  m_{i,j}
  :=
  \delta^{t(j)-1}\prod_{k\prec j}(1-p_i^k)p_i^j
\]
be agent \(i\)'s true discounted reach-and-success mass. With the notation of
Appendix~\ref{app:bcr}, coordinatewise comparison gives
\[
  L_j\le m_{i,j}\le U_j
  \qquad\text{for every }i,j.
  \tag{F.4}
\]
Choose for each \(j\) the upper-endpoint product prior (B.4), so
\(J_j(\nu^j)=U_j\).

Combine agent \(i\)'s success terms on \(j\) into
\[
  e_{i,j}:=
  \begin{cases}
    y_j,       & i\text{ is only the reader}, \\
    v_i^j-y_j, & i\text{ is only the writer}, \\
    v_i^j,     & i\text{ holds both roles},   \\
    0,         & i\text{ holds neither role}.
  \end{cases}
  \tag{F.5}
\]
The no-overbidding argument above handles a writer distinct from the reader, and the
internal-transfer identity handles an agent with both roles. Hence
\[
  0\le e_{i,j}\le v_j^{\max}.
  \tag{F.6}
\]
Let \(t_i\) be agent \(i\)'s total up-front reader payment. Its own-belief utility is
\[
  u_i=\sum_jm_{i,j}e_{i,j}-t_i.
\]
The all-zero report is an opt-out deviation and gives nonnegative utility, so
\(u_i\ge0\). Under the certificates,
\[
  u_i^{\mathrm{cert}}
  =
  \sum_jU_je_{i,j}-t_i
  \ge
  \sum_jm_{i,j}e_{i,j}-t_i
  =u_i
  \ge0.
  \tag{F.7}
\]
Thus the agent-by-agent certificate holds. Writer-to-reader transfers cancel
statewise, retained revenue is the sum of the up-front payments, and aggregate
allocation on box \(j\) is at most \(v_j^{\max}\). Equations (F.4)--(F.7) prove
Theorem~4. Notice that exact writer-allocation Envy-Free is nowhere used.

%% file: appendices/appendix-g-reported-values-mixed-revenue.tex
\appendixsection{Ordinary Mixed Revenue for the Second-Price Mechanism}{app:reported-mixed}

We derive expected-revenue bounds for ordinary mixed equilibria of RWSP that satisfy
exact writer-allocation Envy-Free at almost every
realized report profile. The bounds include a common guarantee under constant
probability competition.

Put
\[
  d:=1-\delta,
  \qquad
  c_\delta
  :=
  \min_{\substack{0<a\le1\\
      a\ge\delta[1-a\log(1/a)]}}
  [1-a\log(1/a)].
  \tag{G.1}
\]
The inequality \(a\log(1/a)\le e^{-1}\) gives
\(c_\delta\ge1-e^{-1}\).

Assume finitely many agents and boxes, \(0<p_i^j<1\), and finite
\(v_i^j\ge0\). Probability and value reports range over \([0,1]\) and
\([0,\infty)\), respectively, so every full-report deviation below is admissible.
Put \(p_j^{\max}:=\max_i p_i^j\), \(v_j^{\max}:=\max_i v_i^j\), let
\(v_j^{(2)}\) be the second-highest true value on box \(j\), and put
\(W^{(2)}:=W_\delta(p^{\max},v^{(2)})\).
Fix an ordinary mixed equilibrium of the mechanism and deterministic role, second-price,
and tie rules in Section~\ref{reported-mechanism}. Assume exact writer-allocation
Envy-Free at almost every realized report profile. Let \(R\) be expected up-front
reader-payment revenue and \(u_i\) the expected utility of agent \(i\). Fix a target
box \(B\) and a set \(H\) of \(K\ge3\) agents. Write
\[
  x_i(B):=z(p_i^B,v_B^{(2)}),
  \qquad
  r_i(B):=p_i^Bv_B^{(2)}.
  \tag{G.2}
\]
Let \(g_i(B)\) be the probability that replacing only agent \(i\)'s value report on
\(B\) by zero lowers the second price below \(v_B^{(2)}\). Pointwise Envy-Free gives
\(y_B\ge v_B^{(2)}\), so at least two realized value reports reach this floor. Therefore
\[
  \sum_{i\in H}g_i(B)\le2.
  \tag{G.3}
\]

Order the \(x_i(B)\)'s within \(H\), and let \(A_x(B,H)\) be their sum after deleting
the two largest. Define \(A_r(B,H)\) analogously. For \(A,U\ge0\), put
\[
  \Psi_\delta(A,U)
  :=
  \sup_{s>0}
  \frac{[(1-e^{-s})A-sU]_+}{1+s\delta}.
  \tag{G.4}
\]

\subsection{The equilibrium-specific bound}

\textbf{Theorem G.1.}
For every \(B\) and \(H\),
\[
  R
  \ge
  \max\left\{
  \frac dK
  \left[
  c_\delta d\sum_{i\in H}(1-g_i(B))x_i(B)
  -\sum_{i\in H}u_i
  \right]_+,
  \right.
\]
\[
  \left.
  d\sup_{(s_i)\in(0,\infty)^H}
  \frac{
  \left[
  \sum_{i\in H}(1-e^{-s_i})(1-g_i(B))r_i(B)
  -\sum_{i\in H}s_i u_i
  \right]_+
  }{
  K+\delta\sum_{i\in H}s_i
  }
  \right\}.
  \tag{M1}
\]
In particular,
\[
  R
  \ge
  \frac dK
  \max\left\{
  \left[c_\delta dA_x(B,H)-\sum_{i\in H}u_i\right]_+,
  \Psi_\delta\left(A_r(B,H),\sum_{i\in H}u_i\right)
  \right\}.
  \tag{M1c}
\]

\emph{Proof.}
Fix \(i\in H\). Replace both reports of \(i\) by zero on every box, and let \(T_i\)
be the largest remaining index. Restoring \(i\)'s reports weakly raises every index.
The first full-profile payment is therefore at least \(dT_i\), so
\[
  \mathbb E[T_i]\le\frac Rd=:C.
  \tag{G.5}
\]
Let \(G_i\) be the event that removal of \(i\)'s target value report preserves the
floor \(v_B^{(2)}\), and put \(g_i=\Pr(G_i^c)\).

For \(0<x<x_i(B)\), let \(q_i(x)\) satisfy
\(z(q_i(x),v_B^{(2)})=x\). Consider the fixed deviation that reports zero everywhere
except for probability \(q_i(x)\) and value zero on \(B\). On
\(G_i\cap\{T_i<x\}\), box \(B\) is uniquely first, \(i\) is its unique reader, and
its target utility is at least
\[
  (p_i^B-q_i(x))v_B^{(2)}
  \ge d(x_i(B)-x).
\]
If the target is later, its payment is at most \(\delta T_i\). The Nash inequality
therefore gives
\[
  u_i+\delta C
  \ge
  d(x_i(B)-x)\Pr(G_i,T_i<x).
  \tag{G.6}
\]
Set \(h_i:=1-g_i\). From
\(\Pr(G_i,T_i<x)\ge\Pr(T_i<x)-g_i\), integrate the implied upper bound on
\(\Pr(T_i<x)\) up to the point where it becomes vacuous. The layer-cake identity and
(G.5) yield
\[
  u_i+C
  \ge
  c_\delta d(1-g_i)x_i(B).
  \tag{G.7}
\]
For completeness, after normalizing
\[
  a:=\frac{u_i+\delta C}{d(1-g_i)x_i(B)},
  \qquad
  r:=\frac{u_i+C}{d(1-g_i)x_i(B)},
\]
the integration gives \(r\ge1-a\log(1/a)\), while \(u_i\ge0\) gives
\(a\ge\delta r\). Definition (G.1) then implies \(r\ge c_\delta\).

The second deviation draws the target probability report \(Q_i\) from
\([0,(1-e^{-s})p_i^B]\) with density
\[
  f_{i,s}(q)=\frac1{s(p_i^B-q)},
  \tag{G.8}
\]
reports value zero on \(B\), and reports zero elsewhere. Conditional on \(G_i\),
integrating the first-position reader margin above the reduced-index threshold gives
\[
  su_i+(1+s\delta)C
  \ge
  (1-e^{-s})(1-g_i)r_i(B).
  \tag{G.9}
\]
A target that cannot become first makes the right side of the underlying pointwise
bound nonpositive, so (G.9) also covers the boundary cases.

Sum (G.7) over \(H\), sum (G.9) using arbitrary \(s_i>0\), and substitute
\(C=R/d\). This proves (M1). Equation (G.3) implies
\[
  \sum_{i\in H}(1-g_i)x_i(B)\ge A_x(B,H),
  \qquad
  \sum_{i\in H}(1-g_i)r_i(B)\ge A_r(B,H).
\]
Using a common \(s_i=s\) and optimizing proves (M1c). \(\square\)

\subsection{The type-dependent bound}

Define
\[
  H_i:=W_\delta(p_i,v^{\max}),
\]
and, for \(1\le L\le M\),
\[
  T_{i,L}^{\mathrm{tail}}
  :=
  \max_\pi
  \sum_{t=L+1}^M
  \delta^{t-1}
  \left(\prod_{r<t}(1-p_i^{\pi_r})\right)
  p_i^{\pi_t}v_{\pi_t}^{\max},
  \qquad
  \Delta_{i,L}:=H_i-T_{i,L}^{\mathrm{tail}}.
  \tag{G.10}
\]
At most \(2L\) distinct reader-or-writer identities occur in the first \(L\)
positions. If \(\Delta_{H,L}^{[\ell]}\) denotes the \(\ell\)-th largest
\(\Delta_{i,L}\) among \(i\in H\), put
\[
  \mathcal U_H(L)
  :=
  \sum_{i\in H}T_{i,L}^{\mathrm{tail}}
  +
  \sum_{\ell=1}^{\min\{2L,K\}}\Delta_{H,L}^{[\ell]},
  \qquad
  \mathcal U_H^*:=\min_{1\le L\le M}\mathcal U_H(L).
  \tag{G.11}
\]
Every combined success value lies in \([0,v_j^{\max}]\) by Envy-Free. If an agent has
no early role, its gross utility is bounded by its exact tail term; otherwise it is
bounded by \(H_i\). Up-front payments are nonnegative. Hence
\[
  \sum_{i\in H}u_i\le\mathcal U_H^*.
  \tag{G.12}
\]
Substitution in (M1c) proves
\[
  R
  \ge
  \frac dK
  \max\left\{
  [c_\delta dA_x(B,H)-\mathcal U_H^*]_+,
  \Psi_\delta(A_r(B,H),\mathcal U_H^*)
  \right\}.
  \tag{M2}
\]

\subsection{Probability-competition consequences}

Assume every box \(B\) has a set \(H_B\) of \(K\ge3\) agents satisfying
\(p_i^B\ge\alpha p_B^{\max}\). Then
\[
  A_x(B,H_B)\ge\alpha(K-2)z(p_B^{\max},v_B^{(2)}),
  \qquad
  A_r(B,H_B)\ge\alpha(K-2)p_B^{\max}v_B^{(2)}.
  \tag{G.13}
\]
Define
\[
  H^{\max}:=\max_iH_i,
  \qquad
  \rho_L:=
  \begin{cases}
    \delta^L, & L<M, \\
    0,        & L=M,
  \end{cases}
  \qquad
  \eta_{K,L}:=\rho_L+(1-\rho_L)\frac{\min\{2L,K\}}K.
  \tag{G.14}
\]
The tail terms in (G.11) give
\[
  \mathcal U_{H_B}^*\le K\eta_{K,L}H^{\max}.
  \tag{G.15}
\]
Also
\[
  \max_Bz(p_B^{\max},v_B^{(2)})\ge W^{(2)},
  \qquad
  \max_Bp_B^{\max}v_B^{(2)}
  \ge\frac{W^{(2)}}{G_M(\delta)}.
  \tag{G.16}
\]
Apply (M2) to every target and use (G.13)--(G.16). This gives the common bound
\[
  R
  \ge
  d\max_{1\le L\le M}
  \max\left\{
  \left[
  c_\delta d\alpha\left(1-\frac2K\right)W^{(2)}
  -\eta_{K,L}H^{\max}
  \right]_+,
  \right.
\]
\[
  \left.
  \Psi_\delta\left(
  \frac{\alpha}{G_M(\delta)}
  \left(1-\frac2K\right)W^{(2)},
  \eta_{K,L}H^{\max}
  \right)
  \right\}.
  \tag{M3}
\]
Equations (M1)--(M3) prove the ordinary mixed-equilibrium revenue bounds. All deviations were chosen independently of
the realized opponents' reports. No value competition or interior equilibrium-report
assumption entered the proof.

%% file: appendices/appendix-h-reported-values-constant-competition.tex
\appendixsection{Constant Probability Competition with Reported Values}{app:reported-constant}

\textbf{Theorem 5} (Reported-values comparison). \emph{For every
  \(0<\alpha,\beta\le1\) and \(0<\delta<1\), there is a finite probability depth
  \(K_0(\alpha,\beta,\delta)\), independent of the number of boxes, such that every
  existing pure equilibrium with interior probability reports that satisfies constant
  probability competition, qualified value competition, and pointwise exact
  writer-allocation Envy-Free satisfies}
\[
  \operatorname{Rev}_{y}
  =\Omega\!\left(
  \alpha\beta(1-\delta)^3
  \operatorname{OPT}_{\mathrm{RV\text{-}DNBR}}
  \right).
  \tag{SV-COMP}
\]

\emph{Proof.}
Fix an equilibrium satisfying pointwise exact writer-allocation Envy-Free and constant
probability competition. For the comparison with dynamic NBR, also assume a pure equilibrium in the
interior probability-report domain of Theorem~4.

Put \(p_j^{\max}:=\max_i p_i^j\), \(v_j^{\max}:=\max_i v_i^j\), let
\(v_j^{(2)}\) be the second-highest true value on box \(j\), and define
\(W^{\max}:=W_\delta(p^{\max},v^{\max})\) and
\(W^{(2)}:=W_\delta(p^{\max},v^{(2)})\).
Assume qualified value competition \(v_j^{(2)}\ge\beta v_j^{\max}\). Benchmark
monotonicity and homogeneity give
\[
  H^{\max}
  =
  \max_iW_\delta(p_i,v^{\max})
  \le W^{\max}
  \le\beta^{-1}W^{(2)}.
  \tag{H.1}
\]
Define
\[
  \theta:=c_\delta(1-\delta)\alpha\beta,
  \qquad
  L_0:=\left\lceil\frac{\log(8/\theta)}{\log(1/\delta)}\right\rceil,
\]
\[
  K_0(\alpha,\beta,\delta)
  :=
  \max\left\{
  4,\left\lceil\frac{16L_0}{\theta}\right\rceil
  \right\}.
  \tag{H.2}
\]
Choose \(K=K_0\) and \(L=\min\{M,L_0\}\) in the first branch of (M3).
Definitions (G.14) and (H.2) give
\[
  \eta_{K,L}\le\frac{\theta}{4},
  \qquad
  1-\frac2K\ge\frac12.
  \tag{H.3}
\]
Substituting (H.1)--(H.3) into (M3) proves
\[
  \operatorname{Rev}_{y}
  \ge
  \frac{c_\delta\alpha(1-\delta)^2}{4}W^{(2)}
  \ge
  \frac{c_\delta\alpha\beta(1-\delta)^2}{4}W^{\max}.
  \tag{H.4}
\]
The same proof applies to a pure equilibrium through its degenerate mixed
representation. Thus (H.4) gives the intermediate revenue lower bound in terms of
\(W^{\max}\) and shows explicitly that \(K_0\) is independent of \(M\).

If the equilibrium is pure and also lies in the interior probability-report domain of
Theorem~4, combine (H.4) with (B.2):
\[
  \operatorname{Rev}_{y}
  \ge
  \frac{c_\delta\alpha\beta(1-\delta)^2}
  {4G_M(\delta)}
  \operatorname{OPT}_{\mathrm{RV\text{-}DNBR}}
  \ge
  \frac{c_\delta\alpha\beta(1-\delta)^3}{4}
  \operatorname{OPT}_{\mathrm{RV\text{-}DNBR}}.
  \tag{H.5}
\]
The first inequality is the sharper finite-\(M\) comparison; the second uses
\(G_M(\delta)\le(1-\delta)^{-1}\). This proves Theorem~5.

%% file: appendices/appendix-i-existence-boundary.tex
\appendixsection{The Repair Potential and the Existence Boundary}{app:existence-boundary}

We show that, under known box values, coordinatewise report-raising repairs weakly
increase RW revenue, strictly increase a repair potential, and converge. Order-changing
deviations prevent this convergence argument from establishing pure-equilibrium
existence. Let \(x=\widehat q\) be the vector of adopted reports. Suppose only coordinate \(j\)
rises from \(x_j\) to \(x_j'>x_j\), and evaluate the new reports in the old order. If \(D_j(x)\) is the old reported discounted reach,
then
\[
  \operatorname{Rev}_{\mathrm{RW}}(x')
  -
  \operatorname{Rev}_{\mathrm{RW}}(x)
  \ge
  (1-\delta)D_j(x)v_j(x_j'-x_j).
  \tag{I.1}
\]
To prove (I.1), let \(W_{j+}(x)\) be the continuation following \(j\). The old-order
change is
\[
  D_j(x)(x_j'-x_j)(v_j-\delta W_{j+}(x)).
\]
Index order gives \(W_{j+}(x)\le\widehat z_j(x)\le v_j\), so the bracket is at least
\((1-\delta)v_j\). Reoptimizing can only raise revenue.

Revenue need not rise strictly when \(D_j(x)=0\). For any \(\eta_q>0\), however,
\[
  \Phi(\widehat q)
  :=
  \operatorname{Rev}_{\mathrm{RW}}(\widehat q)
  +\eta_q\sum_j\widehat q_j
  \tag{I.2}
\]
strictly increases under every nontrivial coordinatewise bid-raising repair. Every
adopted coordinate is nondecreasing and bounded by one, so every repair sequence
converges coordinatewise. RW revenue is the maximum of finitely many continuous
fixed-order objectives and therefore converges as well.

The missing implication is from convergence to Nash equilibrium. A profitable
order-changing deviation may assign the deviator a negative position on the promoted
box while increasing the reach of a pre-existing positive role. No nonnegative
box-level supporter need exist for the promoted report, so the proposed
support-preservation lemma is false. The strict secondary term in (I.2) supplies no
revenue multiplier and does not eliminate simultaneous or order-changing deviations
at the limit. Thus (I.1)--(I.2) prove monotonicity, strict repair potential, and
convergence, but neither existence nor nonexistence of a pure equilibrium.

%% file: appendices/appendix-j-open-domain-nonexistence.tex
\appendixsection{Pure-Equilibrium Nonexistence Below the Unit Report Cap}{app:open-domain-nonexistence}

Consider RW under known box values with probability reports
restricted to \([0,1)\). Every adopted report can then be strictly overtaken.

\phantomsection\label{prop:open-domain-nonexistence}
\textbf{Proposition J.1} (Open-domain nonexistence). \emph{There is a three-box,
eight-agent instance with known box values and a constant \(\varepsilon_0>0\) such that, when
\(q_i^j\in[0,1)\), the RW game has no pure \(\varepsilon\)-equilibrium for any
\(\varepsilon<\varepsilon_0\), under any complete tie rule. Every true success
probability in the instance is at most \(3/5\).}

Set
\[
  \delta=\frac45,
  \qquad
  (v_A,v_B,v_C)=(100,1,1).
  \tag{J.1}
\]
There are four anchor agents and four tail agents, with true success probabilities
\[
  \begin{array}{c|ccc}
        & p^A  & p^B  & p^C   \\ \hline
    H_1 & 0.50 & 0    & 0     \\
    H_2 & 0.50 & 0.01 & 0     \\
    H_3 & 0.50 & 0    & 0.01  \\
    H_4 & 0.50 & 0.01 & 0.01  \\ \hline
    T_1 & 0.20 & 0.50 & 0.60  \\
    T_2 & 0.20 & 0.49 & 0.59  \\
    T_3 & 0.20 & 0.48 & 0.58  \\
    T_4 & 0.20 & 0.47 & 0.57.
  \end{array}
  \tag{J.2}
\]
In particular, \(p_i^j\le 3/5\) for every agent and box.

Fix an arbitrary pure report profile and an arbitrary resolution of all its ties. Write
\(a=\widehat q_A\). Because there are only three reader roles, at least one anchor agent
and at least one tail agent hold no reader role; each such agent has utility zero.

Suppose first that \(a<0.49\). A role-free anchor reports
\[
  q_A'=\frac{a+1/2}{2}>a
\]
on \(A\) and zero on the tail boxes. Since \(q_A'\ge1/4\), box \(A\) is strictly
first. The deviation gives utility
\[
  100\left(\frac12-q_A'\right)
  =50\left(\frac12-a\right)>\frac12.
  \tag{J.3}
\]
Any zero-report tail role assigned to the deviator has zero payment and nonnegative
gross value, so it cannot reduce this gain.

Now suppose that \(a\ge0.49\). Then \(A\) is strictly first, since its reported index
exceeds the largest possible index of either unit-value tail box. Let
\(x=\widehat q_B\) and \(y=\widehat q_C\), and choose a role-free tail agent. Because
\(x,y<1\), that agent can choose
\[
  \max\{x,y\}<r_C<r_B<1,
  \tag{J.4}
\]
report \((r_B,r_C)\) on \((B,C)\), and strictly induce the order \(A,B,C\).
For a tail type \((0.20,p_B,p_C)\), its gross value from the two tail positions converges,
as \(r_B,r_C\uparrow1\), to
\[
  G(p_B,p_C)
  =0.8(1-0.20)p_B
  +0.8^2(1-0.20)(1-p_B)p_C.
  \tag{J.5}
\]
Across the four tail types, the smallest value of \(G\) is \(0.4554752\). The payment
on \(C\) converges to zero, while the payment on \(B\) converges to
\(0.8(1-a)\). Hence the deviator's utility converges to at least
\[
  0.4554752-0.8(1-a)
  \ge 0.4554752-0.8(0.51)
  =0.0474752.
  \tag{J.6}
\]
Thus, for every \(\varepsilon<\varepsilon_0:=0.0474752\), reports satisfying
\((J.4)\) can be chosen close enough to one to give a role-free agent a gain greater
than \(\varepsilon\). The two cases cover every report profile. All relevant takeovers
and box-order comparisons are strict, so the argument is independent of the complete
tie rule. This proves Proposition J.1.

The proposition does not apply to the maintained closed report domain \([0,1]\).
Indeed, the same construction admits a tie-supported pure equilibrium under a suitable
report-blind, agent-specific tie rule when reports equal to one are allowed. Universal
pure-equilibrium existence on \([0,1]\) therefore remains open.

\paragraph{Remark (bidding language).}
This failure may point to a limitation of the bidding language.  The single report
\(q_i^j\) simultaneously describes success and determines the box's contribution to
reported reach.  In the construction, some agents would prefer to express a priority
stronger than any admissible probability report---informally, they would like to report
\(q_i^j>1\).  A richer mechanism might instead ask every agent to report two scalars for
each box: its reach probability and its success probability.  Decoupling these reports
may better represent such preferences and may restore equilibrium in instances of this
kind.  We leave the design and equilibrium analysis of such a mechanism open.